\pdfoutput=1
\documentclass[12pt]{article}

\font\grande=cmr9.5 scaled \magstep4
\font\medio=cmr9.5 scaled \magstep2
\outer\def\beginsection#1\par{\medbreak\bigskip
      \message{#1}\leftline{\bf#1}\nobreak\medskip
\vskip-\parskip
      \noindent}
\usepackage{graphicx} 
\begin{document}
\bibliographystyle {unsrt}

\titlepage
\begin{flushright}
CERN-TH-2018-122
\end{flushright}
\vspace*{1.5cm}

\begin{center}
{\grande Post-inflationary thermal histories}\\
\vspace{6mm}
{\grande and the refractive index of relic gravitons}\\
\vspace{15mm}
 Massimo Giovannini 
 \footnote{Electronic address: massimo.giovannini@cern.ch} \\
\vspace{1cm}
{{\sl Department of Physics, CERN, 1211 Geneva 23, Switzerland }}\\
\vspace{0.5cm}
{{\sl INFN, Section of Milan-Bicocca, 20126 Milan, Italy}}

\vspace*{1cm}
\end{center}

\centerline{\medio  Abstract}
\vspace{5mm}
We investigate the impact of the post-inflationary 
thermal histories on the cosmic graviton spectrum caused by the inflationary variation 
of their refractive index. Depending on the frequency band, the spectral energy distribution 
can be mildly red, blue or even violet. Wide portions of the parameter space lead to 
potentially relevant signals both in the audio range (probed by the advanced generation 
of terrestrial interferometers) and in the mHz band (where space-borne detectors could be 
operational within the incoming score year).  The description of the refractive index in conformally related frames is clarified.

\vskip 0.5cm

\nonumber
\noindent

\vspace{5mm}

\vfill
\newpage

\renewcommand{\theequation}{1.\arabic{equation}}
\setcounter{equation}{0}
\section{Introduction}
\label{sec1}
Stochastic backgrounds of cosmological origin have been suggested 
more than forty years ago \cite{gr1,gr2,par1} as a genuine general relativistic effect 
in curved space-times. Since the evolution of the tensor modes of 
the geometry is not Weyl-invariant \cite{gr1}, the corresponding classical and quantum 
fluctuations  can be amplified not only in anisotropic metric but also in conformally flat 
background geometries \cite{gr2,par1} (see also \cite{par2}). For this reason backgrounds of relic gravitons are
expected, with rather different properties, in a variety of cosmological scenarios and, in particular, 
during an isotropic phase of quasi-de Sitter expansion \cite{star1}. 
The backgrounds of cosmic gravitons are analyzed in terms of the 
spectral energy distribution in critical units, conventionally 
denoted by $\Omega_{gw}(\nu,\tau_{0})$ where $\tau_{0}$ 
is the present value of the conformal time coordinate 
and $\nu$ is the comoving frequency whose numerical value 
coincides (at $\tau_{0}$) with the value of the physical frequency\footnote{In this investigation 
the scale factor is normalized as $a(\tau_{0}) = a_{0} =1$. Natural units $\hbar=c =1$ will be used throughout.}. 
The transition from the radiation-dominated to the matter stage of expansion leads to 
the infrared branch of the spectrum ranging 
between the $\mathrm{aHz}$ and $100 \,\mathrm{aHz}$ \cite{ir1,ir2,ir3}. The standard prefixes
shall be used throughout, i.e.  $1\,\mathrm{aHz} = 10^{-18}, \mathrm{Hz}$, 
$1\, \mathrm{mHz} = 10^{-3}\, \mathrm{Hz}$,  $1\,\mathrm{MHz} = 10^{6} \, \mathrm{Hz}$ 
and so on and so forth. 

Between few aHz and $100$ aHz the low frequency branch of the spectrum 
is universal and it is caused by the tensor modes of the geometry reentering
after matter-radiation equality. For higher frequencies the spectral energy distribution 
bears the mark of the evolution of the Hubble rate prior to the radiation-dominated epoch.
The simplest possibility (so far consistent with observational 
data)  is that  a quasi-de Sitter phase of expansion is followed by a radiation-dominated 
stage: in this case  the spectral energy density is quasi-flat 
\cite{star1,in1,in2,in3} between $100$ aHz and $100$ MHz.
Neglecting all possible complications (damping of the tensor modes due to neutrinos \cite{w1,w2}, 
evolution of relativistic species \cite{l1,l2}, late time dominance of the 
dark energy \cite{l2}) we can estimate\footnote{Note that $h_{0}$ is the present value of the Hubble rate $H_{0}$ in units of $100 \,\mathrm{km}/(\mathrm{sec}\,\times\mathrm{Mpc})$. }
the typical amplitude of the spectral energy distribution in critical units which is 
$h_{0}^{2} \Omega_{gw}(\nu, \tau_{0}) \leq {\mathcal O}(10^{-16.5})$
for frequencies ranging between the mHz and 10 kHz. 
This minute result follows from the  absolute normalization 
of the spectral energy distribution fixed by the upper limit on the tensor to scalar ratio
 $r_{T}(\nu_{p})$ at the pivot frequency\footnote{The scalar and tensor power spectra 
 are customarily assigned at a pivot frequency that is largely conventional. In the present analysis 
 we shall be dealing with a pivot wavenumber   $k_{p} = 0.002 \, \mathrm{Mpc}^{-1}$ corresponding 
to a pivot frequency $\nu_{p}$.} $\nu_{p} = k_{p}/(2\pi) = 3.092\,\,\mathrm{aHz}$.
  
In spite of the fact that the combination of different Cosmic Microwave Background (CMB in what follows) 
observations imply various sets of upper bounds on $r_{T}(\nu_{p})$, here we shall be enforcing the limit 
$r_{T}<0.06$ for the tensor spectral index. This limit follows from a joint analysis of Planck and BICEP2/Keck 
array data \cite{BICPL} (see also \cite{first}). However, in a less conservative perspective we could require that
$r_{T}(\nu_{p}) < 0.17$, as demanded by the WMAP9 results\cite{wmap9a,wmap9b}.
 This particular figure holds if the WMAP9 data are combined with the baryon acoustic oscillation data \cite{bao},  
with the South Pole Telescope data \cite{SPT} and with the Atacama Cosmology Telescope data \cite{ACT}. 
The WMAP9 data (combined with further data sets) lead to bounds on $r_{T}(\nu_{p})$ that are grossly similar, for the 
present ends, to the Planck Explorer data suggesting $r_{T}(\nu_{p}) < 0.11$ \cite{planck}.

One of the tacit (but key) assumptions of the concordance scenario is that radiation 
dominates almost suddenly after the end of inflation. This assumption is used, among other things,
to assess the maximal number of inflationary efolds today accessible by CMB observations.
It is however not unreasonable to presume that in its early stages the Universe passed through different 
rates of expansion deviating from the radiation-dominated evolution.
The slowest possible rate of expansion  occurs when the sound speed of the medium coincides with the 
 speed of light \cite{zel1} (see also \cite{zel2}). Expansion rates even slower than the ones of the stiff phase can only be realized 
 when the sound speed exceeds the speed of light.  This possibility is however not compatible with 
 the standard notion of causality. The plausible range for the existence of such a phase is between the 
 end of inflation and the formation of the light nuclei \cite{mg1,pee,corr}. If the dominance of radiation is to take place already by the 
 time of formation of the baryon asymmetry, then the onset of radiation dominance increases from few MeV to the TeV range. In particular, if the post-inflationary plasma is dominated by a stiff source (i.e. characterized by a barotropic index $ w = p/\rho$ larger than $1/3$) 
 the corresponding spectral energy density inherits a blue (or even violet) slope for typical frequencies larger than the mHz and anyway 
smaller than $100$ GHz. In this case, depending on the parameters characterizing the stiff evolution,
 the spectral energy distribution can be of the order of $10^{-10}$ in the audio band while in the mHz band is at most $10^{-15}$.

In quintessence scenarios the present dominance of a cosmological term is translated into the 
late-time dominance of the potential of a scalar degree of freedom that is called quintessence (see e.g. \cite{wein}).
If we also demand the existence of an early inflationary phase accounting for the existence of large-scale inhomogeneities, the inflaton potential must dominate at early times while the quintessence potential should be relevant much later (see second and third papers in Ref. \cite{mg1} and \cite{pee}).
In between the scalar kinetic term of inflaton/quintessence field dominates the background. When the inflaton and the quintessence field are identified
the existence of this phase is explicitly realized \cite{pee} even if a similar phenomenon may take place also 
in slightly different scenarios. 
  
Gravitational waves might acquire an effective index of refraction when they travel in curved 
space-times \cite{onea,oneb} and this possibility has been recently revisited by 
studying the parametric amplification of the tensor modes 
of the geometry during a quasi-de Sitter stage of expansion \cite{two}:
when the refractive index mildly increases during inflation  
the corresponding speed of propagation of the waves diminishes and 
the power spectra of the relic gravitons are then
blue, i.e. tilted towards high frequencies. The purpose of this paper is 
to compute the spectral energy distribution of the relic gravitons 
produced by a dynamical refractive index without assuming 
a standard post-inflationary thermal history. 

Even though the current upper limits on stochastic backgrounds of relic gravitons 
are still far from their final targets \cite{LV3,LV4}, the advanced Ligo/Virgo projects are described in \cite{LV1,LV2}. 
We shall then suppose, according to Refs. \cite{LV1,LV2}  that the terrestrial interferometers (in their 
advanced version) will be one day able to probe chirp amplitudes ${\mathcal O}(10^{-25})$ corresponding 
to spectral amplitudes $h_{0}^2 \Omega_{gw} = {\mathcal O}(10^{-11})$.  
In the foreseeable future there should be at least one supplementary interferometer operational 
in the audio band namely the Japanese Kamioka Gravitational Wave Detector (for short Kagra) 
\cite{kagra1,kagra2} which is, in some sense, the prosecution and the completion of the 
Tama-300 experiment \cite{TAMA}. In the class of wide-band detectors we should also mention the GEO-600 experiment \cite{GEO1} (which is 
now included in the Ligo/Virgo consortium \cite{GEO2}) and the Einstein telescope \cite{ET1} 
whose sensitivities should definitively improve on the advanced Ligo/Virgo targets.
 
The target sensitivity to detect the stochastic background of inflationary origin 
should correspond to a chirp amplitude $h_{c}={\mathcal O}(10^{-29})$ (or smaller) and 
to a spectral energy distribution in critical units $h_{0}^2 \Omega_{gw} = {\mathcal O}(10^{-16})$ (or smaller). 
These orders of magnitude estimates directly come from the amplitude of the quasi-flat plateau 
produced in the context of single-field inflationary scenarios; in this case the plateau
encompasses the mHz and the audio bands with basically the same amplitude. 
Even though these sensitivities are beyond reach for the current interferometers, 
a number of ambitious projects will be supposedly operational in the future.
The  space-borne interferometers, such as (e)Lisa (Laser Interferometer Space Antenna) 
\cite{LISA}, Bbo (Big Bang Observer) \cite{BBO}, and Decigo 
(Deci-hertz Interferometer Gravitational Wave Observatory) \cite{DECIGO1,DECIGO2}, 
might operate between few mHz and the Hz hopefully within the following score year. 
While the sensitivities of these instruments are still very hypothetical, we can suppose 
(with a certain dose of optimism) that they could even range between 
$h_{0}^2 \Omega_{gw} = {\mathcal O}(10^{-12})$ and $h_{0}^2 \Omega_{gw} = {\mathcal O}(10^{-15})$.

The layout of the paper is the following. In section \ref{sec2} 
the basic action of the problem shall be analyzed in its different forms. For the sake of completeness 
 the relation between different parametrizations will also be discussed with the purpose 
 of arguing that the physical description does not change.  In section \ref{sec3} we shall analyze 
the evolution of the effective horizon and the the amplification of the relic gravitons. 
Section \ref{sec4} will be focussed on the analytic (though approximate) estimates of the graviton spectra while 
section \ref{sec5} contains the discussion of the detectability prospects.
Some concluding remarks are collected in section \ref{sec6}.

\renewcommand{\theequation}{2.\arabic{equation}}
\setcounter{equation}{0}
\section{Gauge-invariance and frame-invariance}
\label{sec2}
Gravitational waves might acquire a refractive index when they evolve in curved space-times \cite{onea,oneb} and 
the impact of this idea on a quasi-de Sitter stage of expansion has been explored in \cite{two} where the presence 
of a (time dependent) refractive index has been introduced for the first time. For standard dispersion relations the propagating 
speed of the tensor modes of the geometry in natural units coincides
with the inverse of the refractive index (i.e. $c_{gw}(\tau) = 1/n(\tau)$)  and the basic 
action can be written, in a covariant language, as:
\begin{equation}
S^{(E)} = \frac{1}{8 \ell_{P}^2} \int d^{4} x \sqrt{-\overline{g}_{(E)}} \biggl[ \overline{g}^{\mu\nu}_{(E)} \partial_{\mu} h^{(E)}_{ij} \partial_{\nu} h^{(E)}_{ij}
+ \biggl(\frac{1}{n^2} -1\biggr) \overline{P}^{\mu\nu}_{(E)} \partial_{\mu} h^{(E)}_{ij} \partial_{\nu} h^{(E)}_{ij} \biggr],
\label{one}
\end{equation}
where  $\ell_{P} = \sqrt{8 \pi G} = 1/\overline{M}_{P}$ and 
$\overline{P}^{\mu\nu}_{(E)}$ is the spatial projector tensor orthogonal to $\overline{u}^{(E)}_{\mu}$: 
\begin{equation}
\overline{P}^{\mu\nu}_{(E)} = \overline{g}^{\mu\nu}_{(E)} - \overline{u}^{\mu}_{(E)} \overline{u}^{\nu}_{(E)}, \qquad 
\overline{g}^{\mu\nu}_{(E)} \overline{u}^{(E)}_{\mu} \overline{u}^{(E)}_{\nu} =1.
\label{onea}
\end{equation}
In the case $n\to 1$ the action of Eq. (\ref{one}) reproduces the original Ford and Paker result \cite{par1}; 
in comoving coordinates\footnote{In the case of a conformally flat metric 
$\overline{g}^{(E)}_{\mu\nu} = a^2_{E} \eta_{\mu\nu}$ (where $\eta_{\mu\nu}$ is the Minkowski metric) we have 
$\overline{u}^{0}_{(E)} = 1/a$.} and in a conformally flat metric Eq. (\ref{one}) assumes the following form:
\begin{equation}
S^{(E)} =  \frac{1}{8 \ell_{P}^2} \int d^{3} x \int d\tau \,\,a^2_{E} \,\,\biggl[ \partial_{\tau} h^{(E)}_{ij} \partial_{\tau} h^{(E)}_{ij} - 
\frac{1}{n^2} \partial_{k} h^{(E)}_{ij} \partial^{k} h^{(E)}_{ij} \biggr].
\label{two}
\end{equation}
The inverse of the refractive index multiplies each spatial derivative 
of the tensor amplitude (see Eq. (\ref{two})). This is the parametrization employed in Refs. \cite{onea,oneb,two} and 
it is physically motivated. It is however possible to adopt a somehow contrived viewpoint and
 to describe the dynamics of the refractive index with an apparently different action namely:
\begin{equation}
S=  \frac{1}{8 \ell_{P}^2} \int d^{3} x \int d\tau \,\,a^2 \,\,n^2 \,\,\biggl[ \partial_{\tau} h_{ij} \partial_{\tau} h_{ij} - \frac{1}{n^2}\partial_{k} h_{ij} \partial^{k} h_{ij} \biggr].
\label{three}
\end{equation}
To get from Eq. (\ref{two}) to Eq. (\ref{three}) we need a specific transformation that leaves unaltered the conformal time 
coordinate and the tensor amplitude while the scale factor is simply rescaled through the refractive 
index itself:
\begin{equation}
a_{E}\to a = \frac{a_{E}}{n}, \qquad h_{ij}^{(E)} \to h_{ij} = h_{ij}^{(E)}.
\label{four}
\end{equation}
This transformation exist and it is nothing but a conformal rescaling.
Indeed Eq. (\ref{four}) transforms separately the background and the tensor inhomogeneities 
but it is not difficult to see that it comes directly from a conformal rescaling 
that leaves unaltered the tensor fluctuations of the geometry.  To make this point more apparent, let us verify explicitly that the transformation (\ref{four}) 
is just a particular case of the following conformal rescaling of the four-dimensional metric:
\begin{equation}
g_{\mu\nu}^{(E)} = \Omega^2 \, G_{\mu\nu}, \qquad \sqrt{-g^{(E)}} = \Omega^4 \, \sqrt{- G}.
\label{five}
\end{equation}
that implies the transformation (\ref{four}) on the background and on the related 
inhomogeneities. From Eq. (\ref{five}) the transformation for the background is immediate and it is given by 
$\overline{g}_{\mu\nu}^{(E)} = \Omega^2 \, \overline{G}_{\mu\nu}$.
In the case of a conformally flat metric of Friedmann-Robertson-Walker type we have $\overline{g}_{\mu\nu}^{(E)} = a^2_{E}(\tau)\, \eta_{\mu\nu}$ and this means
 $\overline{G}_{\mu\nu}^{(E)} = a^2(\tau)\, \eta_{\mu\nu}$. Thus, since $a_{E} = \Omega\, a$ this transformation coincides exactly with the first relation of 
Eq. (\ref{four}) provided, as anticipated, $\Omega \equiv n$. 

The second relation reported in Eq. (\ref{four}) is also a general consequence 
of the conformal rescaling (\ref{five}).  While the proof of this statement is immediate in the case of the tensor modes, 
it is useful to present the complete argument since there are also symmetric implications in the  case of the scalar modes.
 Neglecting, for simplicity, the vector modes of the 
geometry the fluctuations of the metric in the Einstein frame are 
\begin{equation}
g_{\mu\nu}^{(E)}(\vec{x}, \tau) = \overline{g}_{\mu\nu}^{(E)}(\tau) + \delta_{t}g_{\mu\nu}^{(E)}(\vec{x},\tau) 
+ \delta_{s}g_{\mu\nu}^{(E)}(\vec{x},\tau),
\label{seven}
\end{equation}
where  $\delta_{t} g_{\mu\nu}^{(E)}$  and  $\delta_{s} g_{\mu\nu}^{(E)}$ denote respectively the 
tensor and the scalar fluctuations of the geometry in the Einstein frame. In a conformally flat bacgkground 
geometry of Friedmann-Robertson-Walker type
Eq. (\ref{seven}) becomes
\begin{eqnarray}
\delta_{\mathrm{t}} g^{(E)}_{ij}  &=& - a^2_{E}(\tau) \,h_{ij}^{(E)},\qquad \partial_{i} h^{ij}_{(E)} = h_{(E)\,\,i}^{i}=0,
\label{eighta}\\
\delta_{s} g^{(E)}_{00} &=& 2 a_{E}^2 \,\phi_{E},\qquad \delta_{s} g^{(E)}_{i j} 
= 2 a_{E}^2\,( \psi_{E} \delta_{ij} - \partial_{i}\partial_{j } C_{E}),\qquad \delta_{s} g^{(E)}_{0i} = - a_{E}^2 \,\partial_{i} B_{E},
\label{eightb}
\end{eqnarray}
where $h_{ij}^{(E)}$ is the (divergenceless and traceless) tensor amplitude appearing in Eq. (\ref{one}).
By definition the tensor amplitude $h_{ij}^{(E)}$ is invariant under infinitesimal 
diffeomorphisms while the scalar fluctuations are not.

Let us now consider exactly the same decomposition in the conformally related frame defined by Eq. (\ref{five});
as in the case of Eq. (\ref{seven}), the  $G_{\mu\nu}$ can be decomposed into a homogeneous part 
supplemented by its own tensor inhomogeneities: 
\begin{equation}
G_{\mu\nu}(\vec{x}, \tau) = \overline{G}_{\mu\nu}(\tau)+ \delta_{t} G_{\mu\nu}(\vec{x}, \tau) + \delta_{s} G_{\mu\nu}(\vec{x}, \tau),
\label{nine}
\end{equation}
where this time the explicit form of the tensor and scalar fluctuations of the four-dimensional 
metric will be given by:
\begin{eqnarray}
\delta_{t} G_{ij} &=& - a^2(\tau) h_{ij}, \qquad \partial_{i} h^{i}_{j} = h_{i}^{i},
\label{ninea}\\
\delta_{s} G_{00} &=& 2 a^2 \phi,\qquad \delta_{s} G_{i j} 
= 2 a^2 ( \psi \delta_{ij} - \partial_{i}\partial_{j } C),\qquad \delta_{s} G_{0i} = - a^2 \partial_{i} B.
\label{nineb}
\end{eqnarray}
The tensor amplitude $h_{ij}$ defined in Eq. (\ref{ninea}) is gauge-invariant while the scalar 
fluctuations of Eq. (\ref{nineb}) are not immediately gauge-invariant. To work out the relation between the fluctuations 
in the two frames we can therefore start with the tensor modes; from Eq. (\ref{five}), recalling the explicit forms 
of the tensor fluctuations in the two frames (i.e. Eqs. (\ref{eighta}) and (\ref{ninea})) we can  write
\begin{equation}
\delta_{\mathrm{t}} g_{\mu\nu}^{(E)} = \Omega^2 \delta_{t} G_{\mu\nu}, \qquad \overline{g}^{(E)}_{\mu\nu} = \Omega^2(\tau) \overline{G}_{\mu\nu}. 
\label{ten}
\end{equation}
Inserting Eqs. (\ref{eighta}) and (\ref{ninea}) into Eq. (\ref{ten}) we have, as anticipated, that 
\begin{equation}
h_{ij}^{(E)} = h_{ij}, \qquad a_{E}(\tau) = \Omega(\tau) a(\tau).
\label{eleven}
\end{equation}
which coincides with the transformation posited in Eq. (\ref{four}) iff $\Omega= n$.
It is therefore legitimate to conclude that if the two backgrounds are conformally related the gauge-invariant tensor 
amplitudes are also the same in the two frames. In other words the tensor amplitudes 
defined as in Eqs. (\ref{eighta}) and (\ref{ninea}) are both gauge-invariant and frame-invariant. In the conformally related frame the action of Eq. (\ref{one}) becomes
\begin{equation}
S^{(E)} \to S= \frac{1}{8 \ell_{P}^2} \int d^{4} x \biggl\{ \sqrt{-\overline{G}} \biggl[ \overline{G}^{\mu\nu} \Omega^2 \partial_{\mu} h_{ij} \partial_{\nu} h_{ij} + \biggl(\frac{1}{n^2} -1\biggr) \Omega^2 \overline{P}^{\mu\nu}\partial_{\mu} h_{ij} \partial_{\nu} h_{ij} \biggr]\biggr\},
\label{twelve}
\end{equation}
where  the projectors and of the four-velocities have been conformally rescaled as
\begin{equation}
\overline{P}^{\mu\nu}_{(E)} = \frac{1}{\Omega^2} \overline{P}^{\mu\nu}, \qquad \overline{P}^{\mu\nu} = \overline{G}^{\mu\nu} - \overline{U}^{\mu} \overline{U}^{\nu}, \qquad \overline{G}_{\mu\nu} \overline{U}^{\mu} \overline{U}^{\nu} =1,\qquad  \overline{U}^{\mu} = \frac{ \overline{u}^{\mu}_{(E)}}{\Omega}.
\label{thirteen}
\end{equation}
If we now posit that the conformal factor with the refractive index itself coincide (i.e. $\Omega(\tau) = n(\tau)$), the action of Eq. (\ref{thirteen}) becomes exactly, 
\begin{equation}
 S  = \frac{1}{8 \ell_{P}^2}  \int d^{3} x \int d\tau a^2 \biggl[ n^2(\tau) \partial_{\tau} h_{ij} \partial_{\tau} h_{ij} - \partial_{k} h_{ij} \partial_{k} h_{ij} \biggr],\label{fourteena}
\end{equation}
which coincides with the action anticipated in Eq. (\ref{three}). All in all the action of Eqs. (\ref{two}) and (\ref{three}) 
are one and the same action since they are simply related by a conformal rescaling. 

Let us finally mention, as we close the section, the analog results for the scalar modes of the geometry 
which are however less central to the discussion of the present investigation. 
Indeed from Eq. (\ref{five}) we will have that
\begin{equation}
\delta_{s} g_{\mu\nu}^{(E)} = \delta_{s} n \, \overline{G}_{\mu\nu}^{(s)} + n\, \delta_{s} G_{\mu\nu}.
\label{fifteen}
\end{equation}
Recalling then the explicit results of Eqs. (\ref{eightb}) and (\ref{nineb}), 
Eq. (\ref{fifteen}) implies a specific relation between the perturbed 
components in the two frames, i.e. 
\begin{equation}
\phi = \phi_{E} - \frac{1}{2} \biggl(\frac{\delta_{s} n}{n}\biggr),\qquad \psi = \psi_{E} + \frac{1}{2} \biggl(\frac{\delta_{s} n}{n} \biggr), \qquad C = C_{E}, \qquad B = B_{E}.
\label{sixteen}
\end{equation}
Equation (\ref{sixteen}) implies that, unlike their tensor counterparts,
 the scalar inhomogeneities defined in Eqs. (\ref{eightb}) and (\ref{nineb})
are neither gauge-invariant nor frame-invariant.
Note, however, that the curvature perturbations on comoving orthogonal hypersurfaces 
are both gauge-invariant and frame-invariant. Indeed in the two frames they are simply given by 
\begin{equation}
{\mathcal R}_{E} = - \psi_{E} - \frac{{\mathcal H}_{E}}{n^{\prime}} \, \delta_{s} n ,\qquad 
{\mathcal R} = - \psi  - \frac{{\mathcal H}}{n'}\,\delta_{s} n.
\label{seventeen}
\end{equation}
Equation (\ref{seventeen}) does not imply that ${\mathcal R}_{E} \neq {\mathcal R}$, as it could be superficially concluded.
On the contrary, the mismatch between $\psi_{E}$ and $\psi$ is exactly compensated by the mismatch between 
${\mathcal H}_{E}$ and  ${\mathcal H}$. In fact,
from the relation between the background scale factors (i.e. $ a_{E} = a \,n$) we have $2 ({\mathcal H}_{E} - {\mathcal H})= n^{\prime}/n $ so that Eq. (\ref{seventeen}) implies 
${\mathcal R}= {\mathcal R}_{E}$. We therefore have, as anticipated, that the tensor modes of the geometry 
discussed in the bulk of the paper and the curvature perturbations on comoving orthogonal hypersurfaces are both 
frame-invariant and gauge-invariant\footnote{Note that this property has relevant implications in the context 
of some specific class if bouncing models such as the ones proposed in \cite{bb1}.}.

We finally mention that after the appearance of Ref. \cite{two},  two similar papers \cite{three} pursued the same idea. The two approaches ultimately coincide since they are related by a
 conformal rescaling involving the refractive index. More specifically, to get from the description of Ref. \cite{two} to the one of Ref. \cite{three} it is 
 sufficient to make a conformal rescaling and to parametrize the propagating speed or the refractive index 
 as a power of the scale factor. Following the suggestion of Ref. \cite{two}, the authors of Ref. \cite{three} 
considered the evolution of the refractive index in an inflating background. This choice is however 
potentially confusing: since the two descriptions are related by a conformal rescaling
the two backgrounds should also be conformally related \cite{four}. This would mean, in practice, that 
if the background inflates in the Einstein frame, it might not inflate in the 
conformally related frame. However, since the choice of the pivotal frame where the background inflates 
is not constrained, the choice of Ref. \cite{three} is, in a sense, mathematically legitimate but physically 
superficial especially in the light of the previous literature. 
We are therefore in the situation where the two conformally related actions are simply two 
complementary parametrizations of the same effect. To cope with this unwanted 
ambiguity the easiest solution is to define a generalized action for the tensor 
modes encompassing the various possibilities suggested so far. 
As we shall see in sections \ref{sec4} and \ref{sec5} when $\gamma \neq 0$ 
(and, in particular, when $\gamma =1$) the spectral index determined
 in the $\gamma =0$ case is just rescaled by a $\gamma$-dependent prefactor 
 that can be reabsorbed in a redefinition of the spectral index.
 
\renewcommand{\theequation}{3.\arabic{equation}}
\setcounter{equation}{0}
\section{Effective horizons }
\label{sec3}
\subsection{Generalities}
According to the results obtained so far the action describing the evolution of the tensor modes of the geometry in the presence of a dynamical 
refractive index can be parametrized in the following manner:
\begin{equation}
 S = \frac{1}{8 \ell_{P}^2}  \int d^{3} x \int d\tau a^2 n^{2 \gamma} \biggl[\partial_{\tau} h_{ij} \partial_{\tau} h_{ij} - 
 \frac{1}{n^2} \partial_{k} h_{ij} \partial_{k} h_{ij} \biggr].
\label{3firstone}
\end{equation}
When $\gamma=0$ Eq. (\ref{3first}) coincides with the action of Eq. (\ref{one}); conversely if $\gamma =1$ 
the action (\ref{3first}) coincides instead with Eq. (\ref{fourteena}). By keeping the value of $\gamma$ generic the 
two parametrizations can be compared in then light of the present and future detectability prospects. 
It is convenient to simplify the action Eq. (\ref{3first}) by introducing a generalized time coordinate, conventionally denoted by $\eta$:
\begin{equation}
n(\eta) d\eta = d\tau, \qquad b(\eta) = a\, n^{\gamma -1/2}.
\label{etatau}
\end{equation}
 With the redefinition (\ref{etatau}) of the time coordinate,  Eq. (\ref{3firstone}) can be rewritten as 
\begin{equation}
 S = \frac{1}{8 \ell_{P}^2}  \int d^{3} x \int d\eta\, b^2(\eta) \biggl[\partial_{\eta} h_{ij} \partial_{\eta} h_{ij} - 
 \partial_{k} h_{ij} \partial_{k} h_{ij} \biggr].
\label{3first}
\end{equation}
The function $b(\eta)$ plays the role of an effective scale factor: note, in fact, that in the limit $n\to 1$ we have that $\eta$ coincides with $\tau$
 and that, consequently, $b(\eta) \to a(\tau)$. When $n\neq 1$ the evolution of $b(\eta)$ defines 
an effective horizon, namely:
\begin{equation}
{\mathcal F} = \frac{\dot{b}}{b}, \qquad \dot{b} = \frac{\partial b}{\partial \eta} = \frac{1}{n}  \frac{\partial b}{\partial \tau} \equiv \frac{b^{\prime}}{n},
\label{3second}
\end{equation}
where the prime denotes a derivation with respect to the conformal time coordinate $\tau$ while the overdot denotes a derivation with respect to the $\eta$-time (and not a derivation with respect to the cosmic time coordinate, as in the conventional notations). To clarify this point and to avoid potential confusions the following relations are explicitly given:
\begin{eqnarray}
{\mathcal F} &=& \frac{\dot{b}}{b} = \frac{\partial \ln{b}}{\partial \eta} \equiv n a F,\qquad F = \frac{\partial \ln{b}}{\partial t},
\label{3third}\\
{\mathcal H} &=& \frac{a^{\prime}}{a} = \frac{\partial \ln{a}}{\partial \tau} \equiv a H,\qquad H = \frac{\partial \ln{a}}{\partial t},
\label{3fourth}
\end{eqnarray}
which can be verified by using Eq. (\ref{etatau}) and the relation of $\tau$ the cosmic time coordinate $t$, i.e. 
\begin{equation}
n(\eta) d\eta = d\tau = dt /a.
\label{3fifth}
\end{equation}
\subsection{The canonical Hamiltonian and the mode functions}
In terms of the 
canonical normal modes $\mu_{ij}(\vec{x},\eta) = b(\eta) \, h_{ij}(\vec{x},\eta)$ Eq. (\ref{3first}) becomes:
\begin{equation}
S = \frac{1}{8 \ell_{P}^2}  \int d^{3} x \int d\eta\,\biggl[ (\partial_{\eta} \mu_{ij}) (\partial_{\eta} \mu_{ij}) - 2 {\mathcal F}  (\partial_{\eta} \mu_{ij}) \mu_{ij} - 
 (\partial_{k} \mu_{ij}) (\partial_{k} \mu_{ij})\biggr].
 \label{NM1}
\end{equation}
Up to a total time derivative Eq. (\ref{NM1}) can also be written as:
\begin{equation}
S = \frac{1}{8 \ell_{P}^2}  \int d^{3} x \int d\eta\,\biggl[ (\partial_{\eta} \mu_{ij}) (\partial_{\eta} \mu_{ij}) - (\dot{{\mathcal F}} + {\mathcal F}^2)   \mu_{ij} \mu_{ij} - 
 (\partial_{k} \mu_{ij}) (\partial_{k} \mu_{ij})\biggr].
 \label{NM2}
\end{equation}
Since $\mu_{ij}(\vec{x}, \eta)$ is given as the sum over the two polarizations $\oplus$ and $\otimes$
\begin{equation}
\mu_{ij} = \sqrt{2} \, \ell_{P} \sum_{\lambda=\oplus,\otimes} e_{ij}^{(\lambda)} \mu_{\lambda}, \qquad e_{ij}^{(\lambda)} \, e_{ij}^{(\lambda^{\prime})} = 2 \, \delta^{(\lambda\, \lambda^{\prime})},
\label{NM3}
\end{equation}
 the action (\ref{NM2}) becomes immediately: 
\begin{eqnarray}
S &=& \int \, d\eta \, L(\eta),\qquad L(\eta) = \sum_{\lambda=\oplus,\otimes} \int d^{3} x {\mathcal L}_{\lambda}(\vec{x}, \eta),
\label{NM4}\\
{\mathcal L}_{\lambda} &=& \frac{1}{2} \biggl[ \dot{\mu}_{\lambda}^2 - (\dot{{\mathcal F}} + {\mathcal F}^2)   \mu_{\lambda}^2 - (\partial_{k} \mu_{\lambda})^2 \biggr].
\label{NM5}
\end{eqnarray}
From Eqs. (\ref{NM4}) and (\ref{NM5}) the 
 canonical momenta are $\pi_{\lambda} = \dot{\mu}_{\lambda}$; consequently the canonical Hamiltonian associated with 
 Eqs. (\ref{NM4}) and (\ref{NM5}) is given by:
 \begin{equation}
H(\eta) = \sum_{\lambda=\oplus,\otimes} H_{\lambda}(\eta), \qquad H_{\lambda} =\frac{1}{2} \int d^{3} x \, \biggl[ \pi_{\lambda}^2 +  (\dot{{\mathcal F}} + {\mathcal F}^2)  \mu_{\lambda}^2 + 
(\partial_{k} \mu_{\lambda})^2 \biggr].
\label{NM6}
\end{equation}
The commutation relations at equal $\eta$-times 
\begin{equation}
[\hat{\mu}_{\lambda}(\vec{x}, \eta), \, \hat{\pi}_{\lambda^{\prime}}(\vec{y}, \eta) ] = i\, \delta^{(3)}(\vec{x} - \vec{y}) \, \delta_{\lambda\, \lambda^{\prime}},
\label{NM7}
\end{equation}
together with the explicit form of the Hamiltonian (\ref{NM6}) lead directly to the evolution equations of the operators $\hat{\mu}_{\lambda}$ and $\hat{\pi}_{\lambda}$:
\begin{equation}
\partial_{\eta} \hat{\mu}_{\lambda} = \hat{\pi}_{\lambda}, \qquad \partial_{\eta} \hat{\pi}_{\lambda} = ( {\mathcal F}^2 + \dot{{\mathcal F}}) \hat{\mu}_{\lambda} + \nabla^2 \hat{\mu}_{\lambda}.
\label{NM9}
\end{equation}
The Fourier representations of $\hat{\mu}_{\lambda}$ and $\hat{\pi}_{\lambda}$ is:
\begin{eqnarray}
\hat{\mu}_{\lambda}(\vec{x},\eta) = \frac{1}{(2\pi)^{3/2}} \int d^{3} k \biggl[ \hat{a}_{\vec{k},\, \lambda} \, f_{k,\, \lambda} e^{- i \vec{k} \cdot \vec{x}} +  \hat{a}^{\dagger}_{\vec{k},\, \lambda} \, f^{*}_{k,\, \lambda} e^{ i \vec{k} \cdot \vec{x}} \biggr],
\label{NM10}\\
\hat{\pi}_{\lambda}(\vec{x},\eta) = \frac{1}{(2\pi)^{3/2}} \int d^{3} k \biggl[ \hat{a}_{\vec{k},\, \lambda} \, g_{k,\, \lambda} e^{- i \vec{k} \cdot \vec{x}} +  \hat{a}^{\dagger}_{\vec{k},\, \lambda} \, g^{*}_{k,\, \lambda} e^{ i \vec{k} \cdot \vec{x}} \biggr],
\label{NM11}
\end{eqnarray}
where $[ a_{{\vec{k}, \, \lambda}}, \, \hat{a}^{\dagger}_{{\vec{p}, \, \lambda}}] = \delta^{(3)}(\vec{k}-\vec{p}) \, \delta_{\lambda,\, \lambda^{\prime}}$.
The evolution of the mode functions $f_{k,\lambda}$ and $g_{k,\lambda}$  follows from Eq. (\ref{NM9}) while the normalization of their 
Wronskian is a consequence of the commutation relations of Eq. (\ref{NM7}):
\begin{eqnarray}
&& \dot{f}_{k,\,\lambda} = g_{k,\, \lambda}, \qquad \dot{g}_{k,\,\lambda} = - k^2 f_{k,\,\lambda} + (\dot{{\mathcal F}} + {\mathcal F}^2)  f_{k,\,\lambda},
\label{NM12}\\
&& f_{k,\, \lambda}(\eta) f^{*}_{k,\, \lambda}(\eta) - f_{k,\, \lambda}^{*}(\eta) g_{k,\, \lambda}(\eta) = i.
\label{NM13}
\end{eqnarray}
The equations for the mode functions reported in Eq. (\ref{NM12}) can be decoupled as:
\begin{equation}
\ddot{f}_{k}  + \biggl[k^2 - \frac{\ddot{b}}{b} \biggr]f_{k} =0, \qquad g_{k}= \dot{f}_{k},
\label{NM14}
\end{equation}
where the polarization index has been omitted since the result of Eq. (\ref{NM14}) holds both for $\oplus$ and for $\otimes$. By recalling that $\hat{h}_{ij} b = \hat{\mu}_{ij}$ the mode expansion of the tensor amplitude $\hat{h}_{ij}(\vec{x},\eta)$ in the 
$\eta$-time is given by:
\begin{equation}
\hat{h}_{ij}(\vec{x},\eta) = \frac{\sqrt{2} \ell_{P}}{(2\pi)^{3/2} b(\eta)}\sum_{\lambda} \int \, d^{3} k \,\,e^{(\lambda)}_{ij}(\vec{k})\, [ f_{k,\lambda}(\eta) \hat{a}_{\vec{k}\,\lambda } e^{- i \vec{k} \cdot \vec{x}} + f^{*}_{k,\lambda}(\eta) \hat{a}_{\vec{k}\,\lambda }^{\dagger} e^{ i \vec{k} \cdot \vec{x}} ],
\label{NM15}
\end{equation}
where the explicit form of the two polarizations can be written as:
 \begin{equation}
 e_{ij}^{(\oplus)}(\hat{k}) = (\hat{m}_{i} \hat{m}_{j} - \hat{q}_{i} \hat{q}_{j}), \qquad 
 e_{ij}^{(\otimes)}(\hat{k}) = (\hat{m}_{i} \hat{q}_{j} + \hat{q}_{i} \hat{m}_{j}),
 \label{NM16}
 \end{equation}
and  $\hat{k}_{i} = k_{i}/|\vec{k}|$,  $\hat{m}_{i} = m_{i}/|\vec{m}|$ and $\hat{q} =q_{i}/|\vec{q}|$ are three mutually 
orthogonal directions  and $\hat{k}$. If we now represent the field operator $\hat{h}_{ij}(\vec{x},\eta)$ in Fourier space:
\begin{equation}
\hat{h}_{ij}(\vec{p},\eta) = \frac{1}{(2\pi)^{3/2}}\,  \, \int d^{3} x \, \hat{h}_{ij}(\vec{x},\eta)\,\, e^{ i \vec{p}\cdot\vec{x}},
\label{St0C}
\end{equation}
we  also have from Eqs. (\ref{NM15}) and (\ref{St0C}):
\begin{equation}
\hat{h}_{ij}(\vec{p}, \eta) = \frac{1}{b}\sum_{\lambda} \biggl[ e_{ij}^{(\lambda)}(\hat{p}) f_{k,\, \lambda}(\eta) \hat{a}_{\vec{p}\, \lambda} +
e_{ij}^{(\lambda)}(-\hat{p})  f_{k,\, \lambda}^{*}(\eta)  \hat{a}_{-\vec{p}\, \lambda}^{\dagger} \biggr].
\label{TTT1}
\end{equation}
It follows from Eq. (\ref{TTT1}) that the two-point function in real and in Fourier space is given by 
The two-point functions computed from Eq. (\ref{TTT1}) are simply\footnote{For the sake of notational 
accuracy, we remind that, throughout this analysis, natural logarithms will be denoted by $\ln$ while the common logarithms 
will be denoted by $\log$.}
\begin{eqnarray}
\langle \hat{h}_{ij}(\vec{x}, \eta) \, \hat{h}_{ij}(\vec{x}+ \vec{r},\eta) \rangle &=& \int d\ln{k} \, {\mathcal P}_{T}(k,\eta) j_{0}(k r), 
\label{Q12a}\\
\langle \hat{h}_{ij}(\vec{k},\eta) \, \hat{h}_{mn}(\vec{p},\eta) \rangle &=& \frac{2\pi^2}{k^3} {\mathcal P}_{T}(k,\eta) \, {\mathcal S}_{ijmn}(\hat{k}) \delta^{(3)}(\vec{k} +\vec{p}),
\label{Q12b}
\end{eqnarray}
where $j_{0}(kr)$ is the spherical Bessel function of zeroth order \cite{abr1,abr2}. The tensor power spectrum of Eqs. (\ref{Q12a}) and (\ref{Q12b}) is
then given by
\begin{eqnarray}
{\mathcal P}_{T}(k,\eta) &=& \frac{4 \ell_{P}^2}{\pi^2 b^2(\eta)} k^3 |f_{k}(\eta)|^2, 
\label{Q13a}\\
{\mathcal S}_{ijmn}(\hat{k}) &=& \frac{1}{4} \biggl[ p_{mi}(\hat{k}) p_{nj}(\hat{k}) + p_{mj}(\hat{k}) p_{ni}(\hat{k}) - p_{ij}(\hat{k}) p_{m n}(\hat{k}) \biggr]
\nonumber\\
&\equiv& \sum_{\lambda} e_{ij}^{(\lambda)}(\hat{k}) \, e_{m n}^{(\lambda)}(\hat{k})/4, \qquad p_{ij}(\hat{k}) = (\delta_{i j} - \hat{k}_{i} \hat{k}_{j}).
\label{Q13b}
\end{eqnarray} 

\subsection{Evolution of the effective horizon}
The variation of the refractive index can be measured in units of the Hubble rate in full analogy with 
what it is customarily done in the case of the slow-roll parameter, namely:
\begin{equation}
\alpha = \frac{1}{H} \frac{\partial \ln{n}}{\partial t} = \frac{\partial \ln{n}}{\partial\ln{a}}, \qquad \epsilon= - \frac{1}{H^2}\frac{\partial H}{ \,\partial t}.
\label{3sixth}
\end{equation}
Equation (\ref{3sixth}) also implies that the evolution of $n(a)$ could be considered as piecewise 
continuous across a certain critical value of the scale factor $a_{*}$; more specifically the situation 
we are interested in is the one where 
\begin{equation}
n(a) = n_{i} \biggl(\frac{a}{a_{i}}\biggr)^{\alpha}, \qquad a< a_{*},
\label{NNA1}
\end{equation}
while $n(a) = 1$ for $a> a_{*}$. It is relatively simple to imagine a number of continuous interpolation between 
the two regimes but what matters for the present considerations is overall the continuity of $n(a)$, not the specific form
of the profile across the normalcy transition. One of the simplest possibilities is given by\footnote{Note that $n_{i} \geq 1$ but we shall 
always consider the case $n_{i}=1$ as representative of the general situation.}
$n(a,\xi) = n_{i} (a/a_{i})^{\alpha} e^{- \xi a/a_{*}} +1$, going as $a^{\alpha}$ for $a < a_{*}$ 
and approaching $1$ quite rapidly when $a > a_{*}$ and $\xi>1$. The typical scale $a_{*}$ (roughly corresponding to the maximum of $n(a)$) 
may coincide with the end of the inflationary phase but this possibility is neither generic nor compulsory. 
The value of $a_{*}$ corresponds to a critical number of efolds $N_{*}$ 
which is of the order of $N_{t}$ (i.e. the total number of efolds)  if $a_{*}$ marks the end of the 
inflationary phase. This identification is however not mandatory and it will also be 
relevant, from the physical viewpoint,  to consider  the case $N_{*} < N_{t}$ or even $N_{*} \ll N_{t}$. 

It is relevant to mention, for future convenience, that $\dot{b} \geq 0$;
this means that $b(\eta)$ is always an increasing function of the 
$\eta$ coordinate defined in Eq. (\ref{etatau}). 
This observation is important for the forthcoming estimates of the 
cosmic graviton spectrum (see section \ref{sec4} and discussions therein). 
More specifically, if we consider separately the cases
$\gamma =0$ and $\gamma=1$ Eq. (\ref{etatau}) implies: 
\begin{eqnarray}
b(x) &=& \frac{a}{\sqrt{n}} \propto \frac{x}{\sqrt{n_{*} x^{\alpha} e^{- \xi x} +1}}, \qquad \gamma =0,
\label{bonea}\\
b(x) &=& a \sqrt{n} \propto x\,\sqrt{n_{*} x^{\alpha} e^{- \xi x} +1}, \qquad \gamma =1,
\label{btwoa}
\end{eqnarray}
where $n_{*} = n_{i} (a_{*}/a_{i})^{\alpha}$. 
The functions of Eq. (\ref{bonea}) and (\ref{btwoa}) are always 
increasing\footnote{We can take, for instance, $N_{t} = {\mathcal O}(60)$ and 
different values of $N_{*} < N_{t}$. Since $0<\alpha<1$, $b(x)$ increases for $0< x <1$ both in Eqs. (\ref{bonea}) and (\ref{btwoa}).
Moreover $b(x)$ is also increasing for $x>1$. There can be situations 
where, depending on the values of the parameters, the derivative of $b$ with respect to $x$ is always positive except for a small region
 $x = {\mathcal O}(1)$  (i.e. $a \simeq a_{*}$): in these cases the derivative changes sign twice so that $b(x)$ has a local maximum and a local minimum both occurring for  $x = {\mathcal O}(1)$. In spite of that $b(x)$ always increases $x<1$ and for $x>1$.} as a function of $x = a/a_{*}$. Since we shall bound the attention to the case of expanding scale factors,
Eqs. (\ref{bonea})--(\ref{btwoa}) imply that the explicit evolution of $b$ (either in $\tau$ or in $\eta$) is always monotonically increasing.

The relations between $\eta$, $\tau$ and the Hubble radius during the refractive phase are affected by the value of the slow-roll parameter. This is a generic consequence of Eq. (\ref{3fifth})
that fixes the relation between $\eta$ and the conformal time coordinate: 
\begin{equation}
\eta = \int \frac{d \tau}{n} = \int\frac{d a }{a^2 \, H\, n}.
\label{el1}
\end{equation}
If we now integrate Eq. (\ref{el1}) by parts we will have: 
\begin{equation}
\int\frac{d a }{a^2 \, H\, n}  = - \frac{1}{a H n} + \int\frac{d a }{a^2 \, H\, n} (\epsilon -\alpha),
\label{el2}
\end{equation}
implying, together with Eq. (\ref{el1}), that 
\begin{equation}
a \, H \, n = - \frac{1}{(1 - \epsilon + \alpha) \eta}\,.
\label{el3}
\end{equation}
The pump field $\ddot{b}/b$ of Eq. (\ref{NM14}) during the refractive phase
can be written as:
\begin{equation}
\frac{\ddot{b}}{b} = \dot{{\mathcal F}} + {\mathcal F}^2 = n^2 H^2 a^2 [\delta^2 + \delta (1 + \alpha -\epsilon)], \qquad \delta = \alpha (\gamma -1/2) +1.
\label{el4}
\end{equation}
Inerting Eq. (\ref{el3}) into Eq. (\ref{el4}) we finally obtain:
\begin{equation}
 \frac{\ddot{b}}{b} = \frac{\delta^2 +\delta ( 1 -\epsilon +\alpha)}{(1- \epsilon +\alpha)^2 \eta^2},
 \label{el5}
 \end{equation}
 which can also be written as: 
 \begin{equation}
  \frac{\ddot{b}}{b} = \frac{\mu^2 -1/4}{\eta^2}, \qquad \mu = \frac{1}{2} + \frac{\delta}{1 - \epsilon + \alpha}.
 \label{el6}
 \end{equation}
 The same result can be obtained by assuming a slow-roll phase 
 \begin{equation}
 b= a \, n^{\gamma -1/2} = b_{*} \biggl(\frac{a}{a_{*}}\biggr)^{\delta}, \qquad b_{*} = a_{*} n_{*}^{\gamma -1/2},
 \label{el7}
 \end{equation}
 where $a(\tau) = (- \tau/\tau_{*})^{-\beta}$ and $\beta= 1/(1- \epsilon)$. Thus thanks to Eq. (\ref{3fifth}) we have
 \begin{equation}
\biggl(- \frac{\eta}{\eta_{*}}\biggr) = \biggl( - \frac{\tau}{\tau_{*}}\biggr)^{1 + \alpha \beta}, \qquad \eta_{*} = \frac{\tau_{*}}{n_{*}(1 +\alpha \beta)}.
\label{el8}
\end{equation}
The result of Eq. (\ref{el8}) implies:
\begin{equation}
b(\eta) = b_{*} \biggl( - \frac{\eta}{\eta_{*}}\biggr)^{-\nu}, \qquad \nu = \frac{\delta \beta}{1 + \alpha \beta}.
\label{el9}
\end{equation}
If we now compute $\ddot{b}/b$ from Eq. (\ref{el9}) we obtain exactly the result of Eq. (\ref{el6}) where $\mu = (\nu +1/2)$. 
Recalling that $\beta =1/(1-\epsilon)$ we have that $\nu = \delta/(1 -\epsilon + \alpha)$ so that 
the results of Eqs. (\ref{el6}) and (\ref{el9}) coincide and are both consistent with Eq. (\ref{el7}).

\renewcommand{\theequation}{4.\arabic{equation}}
\setcounter{equation}{0}
\section{Cosmic graviton spectra and thermal histories}
\label{sec4}
The cosmic graviton spectra can be estimated analytically \cite{in3,mg1,l1,dd2,absolute} by adapting some of the
standard methods\footnote{These methods must be revisited in a slightly different perspective since the evolution of 
$\eta$ and of the conformal time coordinate only coincide, in the 
present framework, after the end of inflation.} and by noting that Eq. (\ref{NM14}) is equivalent to an integral equation whose initial conditions are assigned at the reference time $\eta_{ex}$:
\begin{eqnarray}
f_{k}(\eta) &=& \frac{b}{b_{ex}} \biggl\{ f_{k}(\eta_{ex}) 
+ \biggl[ \dot{f}_{k}(\eta_{ex}) - {\mathcal F}_{ex} f_{k}(\eta_{ex})\biggr] \int_{\eta_{ex}}^{\eta} \frac{b_{ex}^2}{b^2(\eta_{1})} d\eta_{1}
\nonumber\\
&-&k^2 \int_{\eta_{ex}}^{\eta} \frac{d\eta_{1}}{b^2(\eta_{1})} \int_{\eta_{ex}}^{\eta_{1}}\, b_{ex} \,b(\eta_{2}) \,f_{k}(\eta_{2}) d\eta_{2} \biggr\},
\label{INT1}
\end{eqnarray}
where $\eta_{ex}$ is defined as the turning point at which the solution to Eq. (\ref{NM14}) changes its analytic
form:
\begin{equation}
k^2 = \frac{\ddot{b}_{ex}}{b_{ex}}, \qquad \ddot{b}_{ex} \neq 0.
\label{INT2}
\end{equation}
Equation (\ref{INT2}) can be dubbed by saying that at $\eta_{ex}$ the given mode $k$ exits the effective horizon defined by 
the evolution of $b$; the second requirement of Eq. (\ref{INT2}) is for the moment pleonastic
since the exit always occurs in a regime where $\ddot{b}_{ex} \neq 0$.
Even though $b(\eta)$ never evolves linearly in the vicinity of the exit, this occurrence may arise 
close to the reentry that defines the second relevant turning point of the problem.

\subsection{The large-scale power spectra}
Neglecting the terms ${\mathcal O}(k^2 \eta^2)$ the lowest order solution of Eq. (\ref{INT1}) is:
\begin{eqnarray}
f_{k}(\eta) &=& \frac{b(\eta)}{b_{ex}} \biggl\{ f_{k}(\eta_{ex}) 
+ \biggl[ \dot{f}_{k}(\eta_{ex}) - {\mathcal F}_{ex} f_{k}(\eta_{ex})\biggr] \int_{\eta_{ex}}^{\eta} \frac{b_{ex}^2}{b^2(\eta_{1})} d\eta_{1}\biggr\},
\label{MF3}\\
g_{k}(\eta) &=& \frac{b_{ex}}{b(\eta)} \biggl\{ g_{k}(\eta_{ex}) 
+ \biggl[ \dot{g}_{k}(\eta_{ex}) + {\mathcal F}_{ex} g_{k}(\eta_{ex})\biggr] \int_{\eta_{ex}}^{\eta} \frac{b^2(\eta_{1})}{b_{ex}^2} d\eta_{1}\biggr\},
\label{MF4}
\end{eqnarray}
where, according to Eq. (\ref{NM14}), $\dot{f}_{k}(\eta_{ex}) = g_{k}(\eta_{ex})$ and $\dot{f}_{k}(\eta) = g_{k}(\eta)$. Equations (\ref{MF3}) and (\ref{MF4}) determine the approximate form of the power spectrum for wavelengths larger than the Hubble radius. Since  the second term appearing inside the squared bracket at the right hand side of Eq. (\ref{MF3}) is subleading for typical wavelengths 
larger than the effective horizon, after inserting Eq. (\ref{MF3}) into Eq. (\ref{Q13a}) the tensor power spectrum becomes:
\begin{equation}
{\mathcal P}_{T}(k,\eta) = \frac{2 \, \ell_{P}^2 }{ \pi^2 b_{*}^2 \eta_{*}^2} |A|^2  \, (- k \eta_{*})^{2 ( 1 - \nu)}, \qquad \bigl| A \bigr|= \sqrt{2k} \,\,\bigl| f_{k}(\eta_{ex}) \bigr|,
\label{PS1} 
\end{equation}
where Eq. (\ref{el9}) has been used to get an explicit expression of $b(\eta)$ in the regime $\eta < - \eta_{*}$.
The amplitude $\bigl| A \bigr|$ appearing in Eq. (\ref{PS1}) parametrizes, up to an irrelevant phase, the mismatch between the exact and the approximate 
solutions at $\eta_{ex}$: for $k^2 \ll |\ddot{b}/b|$ the correctly normalized solutions of Eq. (\ref{NM14}) are $f_{k}(\eta) = e^{\pm \,i k\eta}/\sqrt{2 k}$. However as soon as $\eta_{ex}$ is approached the amplitude gets slightly modified and by recalling Eq. (\ref{el6})
the exact solution of Eq. (\ref{NM14}) can be written in terms of Hankel functions \cite{abr1,abr2}
\begin{eqnarray}
f_{k}(\eta) &=& \frac{{\mathcal N}}{\sqrt{2 k}} \sqrt{- k \eta} \, \,H^{(1)}_{\mu}(- k \eta), \qquad {\mathcal N} = \sqrt{\frac{\pi}{2}} e^{i \, \pi(\mu+ 1/2)/2},
\nonumber\\
\mu &=& \nu + \frac{1}{2} =  \frac{3 + 2 \gamma\alpha - \epsilon}{2 ( 1 + \alpha - \epsilon)},
\label{PS2}
\end{eqnarray}
where $H^{(1)}_{\mu}(k \eta)$ is the Hankel function of the first kind\footnote{  Unlike the standard case the argument of the Hankel 
function in Eq. (\ref{PS2}) is not $k \tau$ but rather $k\eta$. Recalling Eq. (\ref{el8}) the solution (\ref{PS2}) is then simple in terms of $\eta$ but not in terms of $\tau$.}. For wavelengths larger than the Hubble radius the Hankel function of Eq. (\ref{PS2}) can be expanded in the limit $|k \eta | \ll 1$ so that thanks to Eq. (\ref{Q13a}) the tensor power spectrum becomes 
\begin{equation}
{\mathcal P}_{T}(k,\eta) = \frac{\ell_{P}^2 2^{ 2 \mu}}{ \pi^3 b_{*}^2 \eta_{*}^2} \Gamma^2(\mu) \, (- k \eta_{*})^{3 - 2\mu}.
\label{PS3} 
\end{equation}
Since $3 - 2 \mu = 2 ( 1 -\nu)$ (as implied by Eq. (\ref{PS2})), the ratio between Eqs. (\ref{PS1}) and 
(\ref{PS3}) implies that:
\begin{equation}
n_{T} = 2 ( 1 -\nu), \qquad \mu = \nu + \frac{1}{2}, \qquad |A(\mu)| = \frac{\Gamma(\mu)}{\sqrt{\pi}} \, 2^{\mu - 1/2},
\label{PS4}
\end{equation}
where $\nu$ has been defined in Eq. (\ref{el9}).
The  value of $|A(\mu)|$ estimates the theoretical error 
of the treatment based on Eq. (\ref{PS1})  and on the approximate form of the mode functions.
While it is often plausible to neglect the complication\footnote{This choice is practical for a swift derivation of the slopes 
characterizing the spectral energy distribution inside the Hubble radius.} of $A(\mu)$ and simply 
set $A(\mu) \to 1$, at low frequencies the absolute normalization of the cosmic graviton spectrum is however very sensitive to the value of the mode functions for  $\eta= {\mathcal O}(\eta_{ex})$. It is then mandatory to use  Eq. (\ref{PS3}) which can also be expressed as:
\begin{equation}
{\mathcal P}_{T}(k,\eta_{*}) = \biggl(\frac{H_{*}}{M_{P}}\biggr)^2\,\, \frac{2^{6 - n_{T}}}{\pi^2} \,\Gamma^2\biggl(\frac{3 - n_{T}}{2}\biggr) \, n_{*}^{ 3 - n_{T} - 2\gamma}\, 
\biggl| 1 + \frac{\alpha}{1 - \epsilon}\biggr|^{2 -n_{T}} \,\,\biggl( \frac{k}{a_{*} H_{*}}\biggr)^{n_{T}},
\label{PS5}
\end{equation}
where  $H_{*}$ denotes the Hubble rate at $\eta_{*}$. 
Equation (\ref{PS5}) is the large-scale power spectrum valid for $k < a_{*} H_{*}$.
The scales that exited the Hubble radius for $\eta > - \eta_{*}$ have a different spectral 
slope and, in this respect, we have a twofold possibility. If $\eta_{*}$ coincides with the end of inflation, 
then the power spectrum will still be given by Eq. (\ref{PS6}) 
where, however, $N_{t} = N_{*}$. Conversely if the refractive phase terminates before the end of inflation 
the power spectrum will have a further branch for $a_{*} H_{*} < k \leq a_{1} H_{1}$:
\begin{equation}
{\mathcal P}_{T}(k,\eta_{*}) = \biggl(\frac{H_{1}}{M_{P}}\biggr)^2\,\, \frac{2^{6 - \overline{n}_{T}}}{\pi^2} \,\Gamma^2\biggl(\frac{3 - \overline{n}_{T}}{2}\biggr)  \,\,\biggl( \frac{k}{a_{1} H_{1}}\biggr)^{\overline{n}_{T}},
\label{PS5a}
\end{equation}
where $\overline{n}_{T} = - 2\epsilon/(1 - \epsilon)$. It is relevant to remark that in the limit $\alpha \to 0 $
we have  $\mu \to (3 - \epsilon)/[2 (1 - \epsilon)]$ where $\mu$ 
is the Bessel index appearing in Eq. (\ref{PS2}). Equation (\ref{PS5a}) can be further modified by appreciating that since between $-\eta_{*}$ and $-\tau_{1}$ the 
background inflates we have 
\begin{equation}
H_{*} a_{*} = ( H_{1} a_{1}) e^{N_{*} - N_{t}}, \qquad n_{*} = n_{i} (a_{*}/a_{i})^{\alpha} \equiv n_{i} e^{\alpha N_{*}}
\label{PS5b}
\end{equation}
Taking into account Eqs. (\ref{PS5a}) and (\ref{PS5b}) the power spectrum (\ref{PS5}) finally becomes:
\begin{eqnarray}
{\mathcal P}_{T}(k,\tau_{1}) &=& \biggl(\frac{H_{1}}{M_{P}}\biggr)^2\,\, q_{T}(n_{i}, N_{t}, N_{*}, n_{T})\,\,
\biggl| 1 + \frac{\alpha}{1 - \epsilon}\biggr|^{2 -n_{T}} \,\,\biggl( \frac{k}{a_{1} H_{1}}\biggr)^{n_{T}},
\label{PS6}\\
q_{T}(n_{i}, N_{t}, N_{*}, n_{T}) &=& \frac{2^{6 - n_{T}}}{\pi^2} \,\Gamma^2\biggl(\frac{3 - n_{T}}{2}\biggr) \, n_{i}^{ 3 - n_{T} - 2\gamma}\, e^{ \alpha N_{*}( 3 - 2 \gamma - n_{T}) - n_{T} (N_{*} - N_{t})},
\end{eqnarray}
where $M_{P} = \sqrt{8 \pi} \,\,\overline{M}_{P}$ (see also the definitions after Eq. (\ref{one})).
Equation (\ref{PS6}) determines the tensor to scalar ratio whose explicit form is
\begin{equation}
r_{T}(k) = \frac{\epsilon}{\pi}
q_{T}(n_{i}, N_{t}, N_{*}, n_{T})\,\,\biggl| 1 + \frac{\alpha}{1 - \epsilon}\biggr|^{2 -n_{T}} \,\,\biggl( \frac{k}{k_{max}}\biggr)^{n_{T}},
\label{PS7}
\end{equation}
where we defined, for the sake of conciseness, $k_{max} = a_{1} H_{1}$. 

\subsection{The power spectra after reentry}
Terrestrial interferometers and space-borne detectors operate 
at the present time and will necessarily measure the cosmic graviton 
spectrum for typical wavelengths shorter than the Hubble radius. 
While the largest wavelengths of the problem (i.e. smallest $k$-modes) 
reentered after matter-radiation equality the shortest wavelengths
(i.e. largest $k$-modes) crossed the effective horizon at 
different epochs after the end of the inflationary stage of expansion.
The reentry depends on the post-inflationary thermal history and on the expansion rate that can be very different from
the one of a radiation-dominated plasma. When the refractive index is not dynamical 
the previous observation leads to a characteristic class of violet spectral energy distribution \cite{mg1,corr,absolute}
and it will be interesting to see what happens in the present situation.
Provided the reentry occurs when:
\begin{equation}
k^2 = \biggl| \frac{\ddot{b}_{re}}{b_{re}} \biggr| , \qquad \ddot{b}_{re} \neq 0,
\label{INT1a}
\end{equation}
then $k \eta_{re} = {\mathcal O}(1)$. However, as already remarked above (see Eq. (\ref{INT2})),  
if $\ddot{b}_{re} \to 0$ in the vicinity of the turning point, then $k \eta_{re} \ll 1$. For $\eta \geq \eta_{re}$ the solution of 
Eq. (\ref{NM14}) can be expressed as:
\begin{equation}
f_{k}(\eta) = c_{+}(k) \overline{f}_{re}(\eta) + c_{-}(k) \overline{f}_{re}^{*}(\eta), \qquad g_{k}(\eta) = c_{+}(k) \overline{g}_{re}(\eta) + c_{-}(k)\overline{g}_{re}^{*}(\eta), 
\label{MF5a}
\end{equation}
where $\overline{f}_{re}(\eta)$ and $\overline{g}_{re}(\eta)$ are the mode functions inside the effective horizon (i.e. quantum mechanically 
normalized plane waves in the crudest approximation). 
From the continuity of $f_{k}(\eta)$ and $g_{k}(\eta)$,  Eqs. (\ref{MF3}) and (\ref{MF4}) imply:
\begin{eqnarray}
f_{k}(\eta_{re}) &=&  \overline{f}_{ex} \biggl( \frac{b_{re}}{b_{ex}}\biggr) + b_{re} b_{ex} (\overline{g}_{ex} - {\mathcal F}_{ex} \overline{f}_{ex}) {\mathcal J}(\eta_{ex}, \eta_{re}), 
\label{MF5}\\
g_{k}(\eta_{re}) &=& \frac{b_{ex}}{b_{re}} \overline{g}_{ex} +  \overline{f}_{ex} \biggl[ \biggl(\frac{b_{re}}{b_{ex}}\biggr) {\mathcal F}_{re} 
- \biggl(\frac{b_{ex}}{b_{re}}\biggr) {\mathcal F}_{ex} \biggr]
\nonumber\\
&+& b_{re} b_{ex} {\mathcal F}_{re} (\overline{g}_{ex} - {\mathcal F}_{ex} \overline{f}_{ex})  {\mathcal J}(\eta_{ex}, \eta_{re}), 
\label{MF6}\\
 {\mathcal J}(\eta_{ex}, \eta_{re}) &=& \int_{\eta_{ex}}^{\eta_{re}} \frac{d\eta}{b^2(\eta)}.
\end{eqnarray}
By continuity Eq. (\ref{MF5a}) evaluated at $\eta_{re}$ must coincide with Eqs. (\ref{MF5})--(\ref{MF6}); thus
$c_{\pm}(k)$ can be determined after some simple algebraic manipulation\footnote{ For the derivation of Eqs. (\ref{MF7})--(\ref{MF8})
it is useful to recall that $\overline{f}_{re} \overline{g}_{re}^{*} - \overline{f}_{re}^{*} \overline{g}_{re} = i$, as implied 
by the constancy of the Wronskian when the second-order differential equation for $f_{k}$ is written in the form (\ref{NM14}); the constancy of the Wronskian also implies for  $\eta > \eta_{re}$ that the coefficients $c_{\pm}(k)$ must satisfy $|c_{+}(k)|^2 - |c_{-}|^2 =1$. }:
\begin{eqnarray}
c_{+}(k) &=& (-i)\biggl\{ \biggl[\overline{g}_{re}^{*} \overline{f}_{ex} \biggl(\frac{b_{re}}{b_{ex}}\biggr) - 
\overline{f}_{re}^{*} \overline{g}_{ex} \biggl(\frac{b_{ex}}{b_{re}}\biggr)\biggr]
- \overline{f}_{re}^{*} \overline{f}_{ex} \biggl[ {\mathcal F}_{re} \biggl(\frac{b_{re}}{b_{ex}}\biggr) - {\mathcal F}_{ex} \biggl(\frac{b_{ex}}{b_{re}}\biggr)\biggr] 
\nonumber\\
&+& b_{re} b_{ex} (\overline{g}_{ex} - {\mathcal F}_{ex} \overline{f}_{ex})  (\overline{g}_{re}^{*} - {\mathcal F}_{re} \overline{f}_{re}^{*}) 
 {\mathcal J}(\eta_{ex}, \eta_{re})\biggr\},
 \label{MF7}\\
 c_{-}(k) &=& i \biggl\{ \biggl[\overline{g}_{re} \overline{f}_{ex} \biggl(\frac{b_{re}}{b_{ex}}\biggr) - \overline{f}_{re} \overline{g}_{ex} \biggl(\frac{b_{ex}}{b_{re}}\biggr)\biggr] - \overline{f}_{re} \overline{f}_{ex} \biggl[ {\mathcal F}_{re} \biggl(\frac{b_{re}}{b_{ex}}\biggr) - {\mathcal F}_{ex} \biggl(\frac{b_{ex}}{b_{re}}\biggr)\biggr] 
\nonumber\\
&+& b_{re} b_{ex} (\overline{g}_{ex} - {\mathcal F}_{ex} \overline{f}_{ex})  (\overline{g}_{re}- {\mathcal F}_{re} \overline{f}_{re}) 
 {\mathcal J}(\eta_{ex}, \eta_{re})\biggr\}.
 \label{MF8}
 \end{eqnarray}
Inside the Hubble radius the mode functions are plane waves or, more precisely, the limit of Hankel functions 
for large values of their arguments \cite{abr1,abr2}. We can then express directly Eqs. (\ref{MF7}) and (\ref{MF8}) bu using the plane wave limit of 
the correspoding mode functions: 
\begin{eqnarray}
c_{+}(k) &=& \frac{ e^{- i k (\eta_{re} - \eta_{ex})}}{2 i k} \biggl[ \frac{b_{re}}{b_{ex}} ( i k - {\mathcal F}_{re}) + 
\frac{b_{ex}}{b_{re}}( ik + {\mathcal F}_{ex})  
\nonumber\\
&+& b_{re} \, b_{ex} ({\mathcal F}_{re} - ik) ({\mathcal F}_{ex} + i k)  {\mathcal J}(\eta_{ex}, \eta_{re})\biggr],
\label{MF11}\\
c_{-}(k) &=& \frac{ e^{- i k (\eta_{re} + \eta_{ex})}}{2 i k} \biggl[ \frac{b_{re}}{b_{ex}} ( i k + {\mathcal F}_{re}) -
\frac{b_{ex}}{b_{re}}( ik + {\mathcal F}_{ex})  
\nonumber\\
&-& b_{re} \, b_{ex} ({\mathcal F}_{re} + ik) ({\mathcal F}_{ex} + i k)  {\mathcal J}(\eta_{ex}, \eta_{re})\biggr].
\label{MF12}
\end{eqnarray}
Since the coefficients $c_{\pm}(k)$ satisfy $|c_{+}(k)|^2 - |c_{-}|^2 =1$ it is sufficient to determine just one of the two square moduli. 
If the exit occurs for $\eta < -\eta_{*}$ and the reentry takes place when the refractive index is not 
dynamical, Eqs. (\ref{MF11}) and (\ref{MF12}) can be 
written more explicitly 
\begin{eqnarray}
c_{+}(k) &=& \frac{ e^{- i k (\tau_{re} - \eta_{ex})}}{2 i k} \biggl[ \frac{a_{re}}{b_{ex}} ( i k - {\mathcal H}_{re}) + 
\frac{b_{ex}}{a_{re}}( ik + {\mathcal F}_{ex})  
\nonumber\\
&+& a_{re} \, b_{ex} ({\mathcal H}_{re} - ik) ({\mathcal F}_{ex} + i k)  {\mathcal J}(\eta_{ex}, \tau_{re})\biggr],
\label{MF11a}\\
c_{-}(k) &=& \frac{ e^{- i k (\tau_{re} + \eta_{ex})}}{2 i k} \biggl[ \frac{a_{re}}{b_{ex}} ( i k + {\mathcal H}_{re}) -
\frac{b_{ex}}{b_{re}}( ik + {\mathcal F}_{ex})  
\nonumber\\
&-& a_{re} \, b_{ex} ({\mathcal H}_{re} + ik) ({\mathcal F}_{ex} + i k)  {\mathcal J}(\eta_{ex}, \tau_{re})\biggr],
\label{MF12a}
\end{eqnarray}
where this time 
\begin{equation}
 {\mathcal J}(\eta_{ex}, \tau_{re}) = \int_{\eta_{ex}}^{\eta_{*}} \frac{d\eta}{b^2(\eta)} +  \int_{\tau_{*}}^{\tau_{re}} \frac{d\tau}{a^2(\tau)}.
 \label{MF13a}
\end{equation}

Because $b(\eta)$ always increases throughout the refractive phase and even later (see Eqs. (\ref{bonea})--(\ref{btwoa}) and discussion therein), 
in Eqs. (\ref{MF11}) and (\ref{MF12})the terms proportional to $|b_{ex}/b_{re}|$ can be neglected in comparison with $|b_{re}/b_{ex}|$. Following this logic the  
approximate form of $|c_{-}(k)|^2$ becomes:
\begin{equation}
|c_{-}(k)|^2 \simeq \frac{1}{4} \biggl(\frac{b_{re}}{b_{ex}}\biggr)^2 \biggl(1 + \frac{{\mathcal F}_{re}^2}{k^2} \biggr) \biggl[ 1 - 2 {\mathcal F}_{ex} b_{ex}^2 {\mathcal J} 
+ b_{ex}^4 ( {\mathcal F}_{ex}^2 + k^2) {\mathcal J}^2\biggr].
\label{sq1}
\end{equation}
Equation (\ref{sq1}) allows for a swift determination of the power spectrum and of the 
spectral energy distribution in the limit $k\tau \gg 1$, i.e. when the relevant wavelengths are all inside the 
Hubble radius:
\begin{eqnarray}
{\mathcal P}(k,\tau) &=& \frac{4 k^2}{\pi^2 \overline{M}_{P}^2} \bigl| c_{-}(k)\bigr|^2 \bigl[ 1 + {\mathcal O} \biggl(\frac{1}{k^2 \tau^2}\biggr) \biggr], 
\label{sq2}\\
\Omega_{gw}(k,\tau) &=& \frac{k^4}{3 H^2 \overline{M}_{P}^2 \pi^2 a^4} \bigl| c_{-}(k)\bigr|^2 \bigl[ 1 + {\mathcal O} \biggl(\frac{1}{k^2 \tau^2}\biggr) \biggr], 
\label{sq3}
\end{eqnarray} 
By taking the ratio between Eqs. (\ref{sq2}) and (\ref{sq3}) we recover the standard relation 
between the power spectrum and the spectral energy density valid when the relevant wavelengths are shorter than the 
Hubble radius at a given epoch:
\begin{equation}
\Omega_{gw}(k,\tau) = \frac{k^2}{12 a^2 H^2} {\mathcal P}_{T}(k,\tau) \biggl[ 1 + {\mathcal O} \biggl(\frac{1}{k^2 \tau^2}\biggr) \biggr].
\label{sq4}
\end{equation}
Consequently inside the Hubble radius we can evaluate indifferently either the power spectrum or the spectral energy distribution.

\subsection{Different thermal histories}
\begin{figure}[!ht]
\centering
\includegraphics[height=7cm]{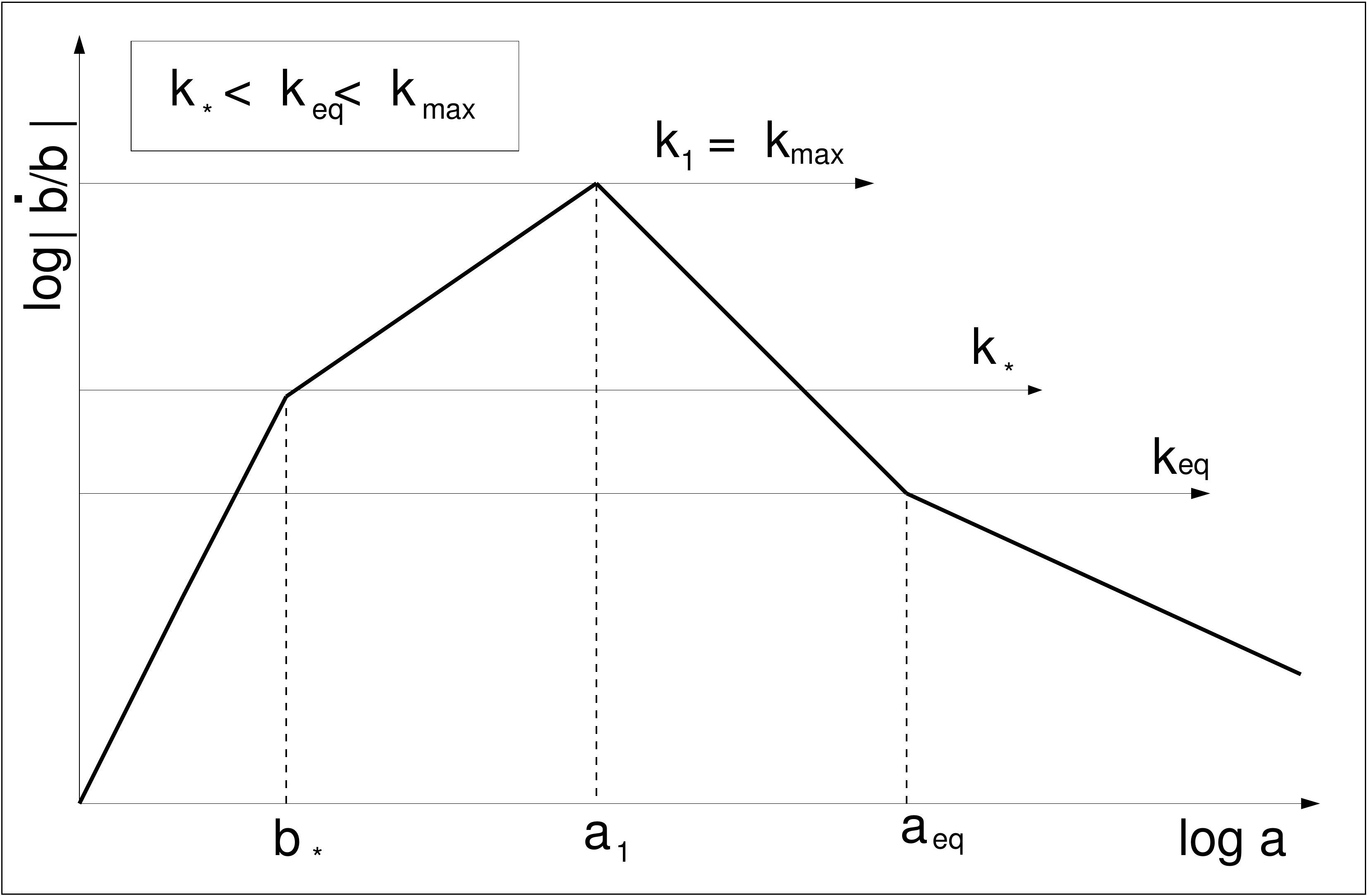}
\caption[a]{We schematically illustrate the evolution of the effective horizon in the minimal 
situation where the radiation background suddenly dominates after inflation.}
\label{FIGU1}      
\end{figure}

The different thermal histories and their effects 
on the spectral energy distribution can be understood by drawing  the salient features of the
effective horizon in various physical situations. In Fig. \ref{FIGU1} on the vertical axis we plot the common logarithm 
of ${\mathcal F} = \dot{b}/b$ and we simultaneously compare it with 
the wavenumbers of the problem\footnote{ This comparison is physically motivated since the crossing condition can also be written as:
$k^2 = {\mathcal F}^2 + \dot{{\mathcal F}}$.}.  In practice the conditions $k\eta_{ex} = {\mathcal O}(1)$ and $k \tau_{re} = {\mathcal O}(1)$ 
will always be verified except that in the case of a reentry during radiation when
$k \tau_{re} \ll 1$ and $a^{\prime\prime}_{re} =0$. 
According to Fig. \ref{FIGU1} we have three different classes of modes: {\it i)} the modes 
exiting the effective horizon during the refractive phase and reentering after equality (i.e. $k < a_{eq} H_{eq}$); {\it ii)} the modes exiting the effective horizon during the refractive phase and reentering 
during radiation (i.e. $a_{eq} H_{eq}< k < a_{*} H_{*}$); {\it iii)} the modes exiting the 
effective horizon after the end of the refractive phase and reentering during radiation 
(i.e. $a_{*} H_{*} < k < a_{1} H_{1}$). The different regions are separated in Fig. \ref{FIGU1}
by three horizontal arrows and the typical wavenumbers (i. e. $k_{1}$, $k_{*}$ and $k_{eq}$) define the three branches 
of the spectral energy density (or of the power spectrum).
Either the inflationary phase continues after $b_{*}$ or the radiation-dominated epoch suddenly kicks in. Between these 
two possibilities the former is more generic than the latter which would correspond, in the 
notation of Fig. \ref{FIGU1}, to the limit $ b_{*} \to a_{1}$; this is why, in Fig. \ref{FIGU1}, we preferred to distinguish 
clearly the two scales by assuming $b_{*} \ll a_{1}$. The three different branches
of the spectral energy distribution illustrated in Fig. \ref{FIGU1} can be deduced from Eqs. (\ref{sq1}), (\ref{sq2}) and (\ref{sq3});
the result of this computation is:
\begin{eqnarray}
\Omega_{gw}(k, \tau_{re}) &\simeq& \biggl(\frac{H_{1}}{M_{P}}\biggr)^2 {\mathcal B}(b_{*}, n_{*}, \epsilon, n_{T})\, \biggl(\frac{k}{a_{*} H_{*}} \biggr) ^{- 2\epsilon/(1 - \epsilon)},\qquad 
a_{*} H_{*} < k \leq a_{1} H_{1},
\label{sq5}\\
\Omega_{gw}(k, \tau_{re}) &\simeq& \biggl(\frac{H_{1}}{M_{P}}\biggr)^2 \, {\mathcal B}(b_{*}, n_{*}, \epsilon, n_{T})\, \biggl(\frac{k}{a_{*} H_{*}} \biggr) ^{n_{T}},\qquad 
a_{eq} H_{eq} < k \leq a_{*} H_{*},
\label{sq6}\\
\Omega_{gw}(k, \tau_{re}) &\simeq& \biggl(\frac{H_{1}}{M_{P}}\biggr)^2  \, {\mathcal B}(b_{*}, n_{*}, \epsilon, n_{T}) \, \biggl(\frac{k}{a_{*} H_{*}} \biggr) ^{n_{T}} \biggl(\frac{k}{a_{eq} H_{eq}} \biggr) ^{-2},\qquad  k \leq a_{eq} H_{eq},
\label{sq7}
\end{eqnarray}
where ${\mathcal B}(b_{*}, n_{*}, \epsilon, n_{T})$  is given by\footnote{Note that 
$\Omega_{M0}$ and $\Omega_{R0}$ denote throughout the 
present values of the critical fractions of matter and radiation in the concordance paradigm.} 
\begin{equation}
{\mathcal B}(b_{*}, n_{*}, \epsilon, n_{T}) = \frac{2}{3\pi} \biggl(\frac{\Omega_{R0}}{\Omega_{M0}}\biggr) n_{*}^{3 - 2 \gamma - n_{T}}\biggl| 1 +\frac{\alpha}{1 - \epsilon} \biggr|^{2 - n_{T}}.
\label{sq8}
\end{equation}
The spectral index $n_{T}$ appearing in Eqs. (\ref{sq5})--(\ref{sq8}) is instead:
\begin{equation}
n_{T} = \frac{(3 - 2 \gamma) \alpha - 2\epsilon}{(1 + \alpha - \epsilon)}=
\frac{\alpha ( 3 - 2 \gamma)}{1 + \alpha} + \frac{[ -2 + \alpha ( 1 - 2\gamma)]\epsilon}{(1 + \alpha)^2} + {\mathcal O}(\epsilon^2),
\label{sq8a}
\end{equation}
where the second equality follows in the limit $\epsilon \ll 1$. As it must, the exact expression of Eq. (\ref{sq8a}) coincides with Eq. (\ref{PS6}).
The quasi-flat branch of Eq. (\ref{sq5}) is caused by the modes that exited the effective horizon for $a > b_{*}$ and reentered during 
the radiation-dominated epoch (i.e. for $a > a_{1}$).  The second branch of the spectrum, reported in Eq. (\ref{sq6}) 
involves the modes that exited the effective horizon during the refractive phase and reentered all along the radiation stage.
Finally the standard infrared branch corresponds to modes that exiting the effective horizon during the refractive epoch and reentering during the matter-dominated phase (i.e. $k < k_{eq}$ in the terminology of Fig. \ref{FIGU1}). 
\begin{figure}[!ht]
\centering
\includegraphics[height=7.5cm]{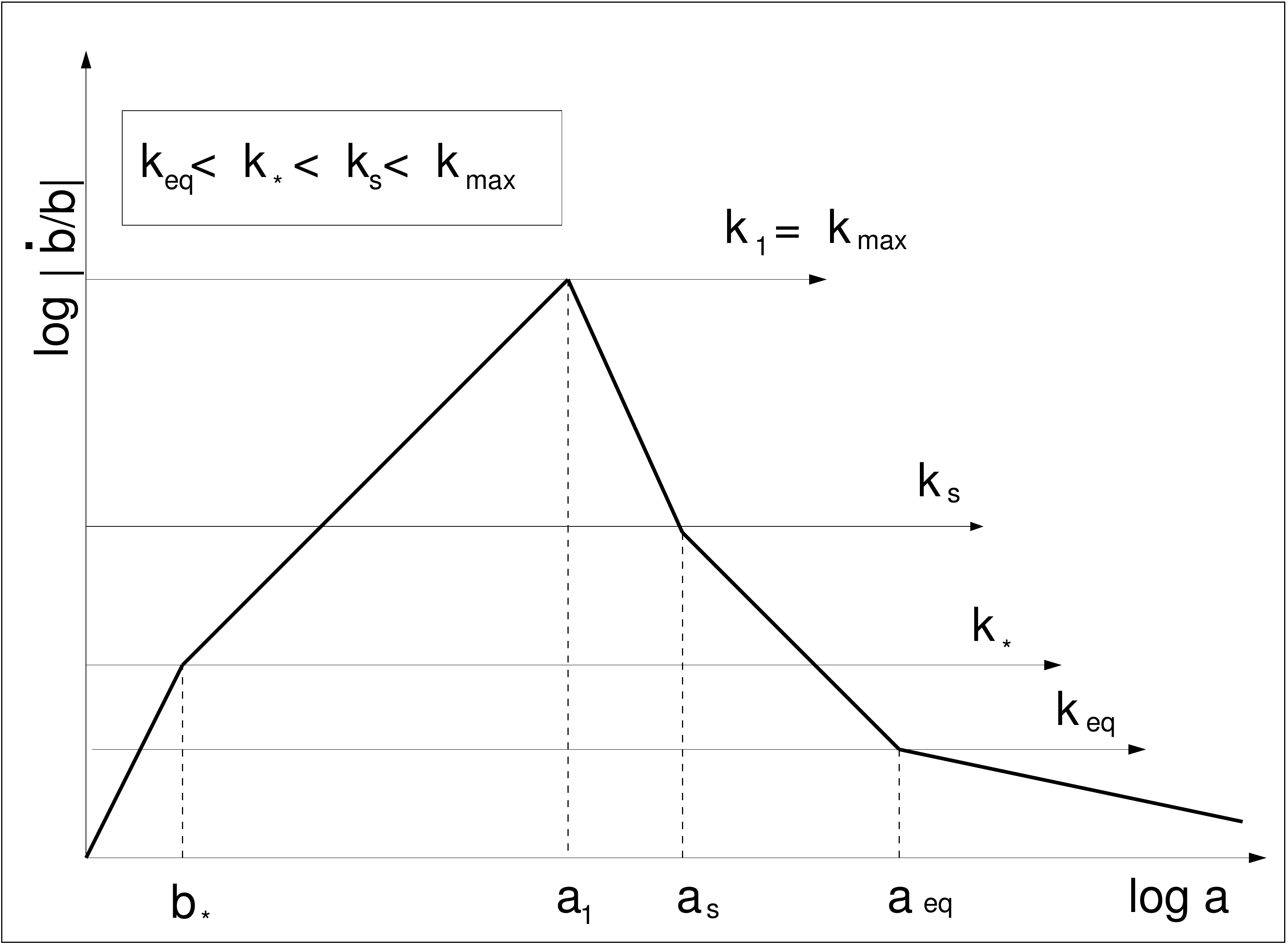}
\caption[a]{We schematically illustrate the evolution of the effective horizon in the case 
 where the dominance of radiation is delayed by the presence of a stiff phase.}
\label{FIGU2}      
\end{figure}
There are no compelling reasons why the physical situation illustrated by Fig. \ref{FIGU1} should be 
considered preferable to some other potentially viable evolution of the effective horizon.
Different thermal histories can be envisaged and cannot be ruled out
by present version of the concordance scenario. Prior to nucleosynthesis, there are no direct 
tests of the thermodynamical state of the Universe and, therefore, the effective equation of state of the primeval plasma 
can be arbitrarily different than the one of radiation.
In Fig. \ref{FIGU2} the effective horizon is illustrated in the case where 
the post-inflationary expansion rate is slower than the one of a radiation-dominated 
plasma; in this case between $a_{1}$ and $a_{s}$ we have
\begin{equation}
\frac{\dot{b}}{b} = {\mathcal H} = a H \propto a^{-(3w +1)/2}, \qquad w > \frac{1}{3}.
\label{sq8b}
\end{equation}
where  $w$ denotes the barotropic index of the stiff post-inflationary phase. 
Note, for comparison, that ${\mathcal H} \propto a$ during inflation while, in the radiation stage, ${\mathcal H} \propto 1/a$
(see aslo Fig. \ref{FIGU1}). In the case of a stiff post-inflationary phase we have instead
that (at most) ${\mathcal H} \propto a^{-2}$, as implied by Eq. (\ref{sq8b}) for $w \to1$.

The evolution sketched in Fig. \ref{FIGU2} leads to a spectral energy distribution characterized by
 four different branches: the wavenumbers $ k < a_{eq} H_{eq}$  correspond 
to scales hitting the effective horizon the first time during the refractive phase and reentering after matter radiation equality: their 
spectral energy distribution will then have the same slope of Eq. (\ref{sq7}). Following the same way of reasoning
 $\Omega_{gw} \propto |k\tau_{*}|^{n_{T}}$ whenever $ a_{eq} H_{eq} < k < a_{*} H_{*}$:
in Fig. \ref{FIGU2} this part of the spectrum corresponds to those modes exiting 
during the refractive phase and reentering during the radiation epoch. 
The supplementary branch of the spectrum implied by Fig. \ref{FIGU2} is caused by 
those modes exiting in the course of the inflationary phase 
and reentering during the stiff phase: in this branch the spectral energy density 
scales as $\Omega_{gw} \propto |k\tau_{s}|^{m_{T}}$ where the spectral index $m_{T}$ 
is now given by:
\begin{equation}
m_{T} = 4 - \frac{2}{1- \epsilon} - \frac{4}{3w +1}.
\label{sq9}
\end{equation}
If the expansion rate 
is slower than radiation the slope in this branch can be very steep (i.e. even violet) with $m_{T} = {\mathcal O}(1)$ in the limit $w \to 1$. 
Incidentally if the expansion rate is faster than radiation\footnote{For instance in the case $w\to 0$ we would have $m_{T} \to -2$.}
it can happen that $m_{T}< 0$.

While the cases illustrated by Figs. \ref{FIGU1} and \ref{FIGU2} 
are the most promising from the viewpoint of the potential signals (as we shall see in the following section), there are other possible
evolutions of the effective horizon 
where the resulting spectral energy distribution does not have a flat (or decreasing) plateau and it always increases.
\begin{figure}[!ht]
\centering
\includegraphics[height=7.5cm]{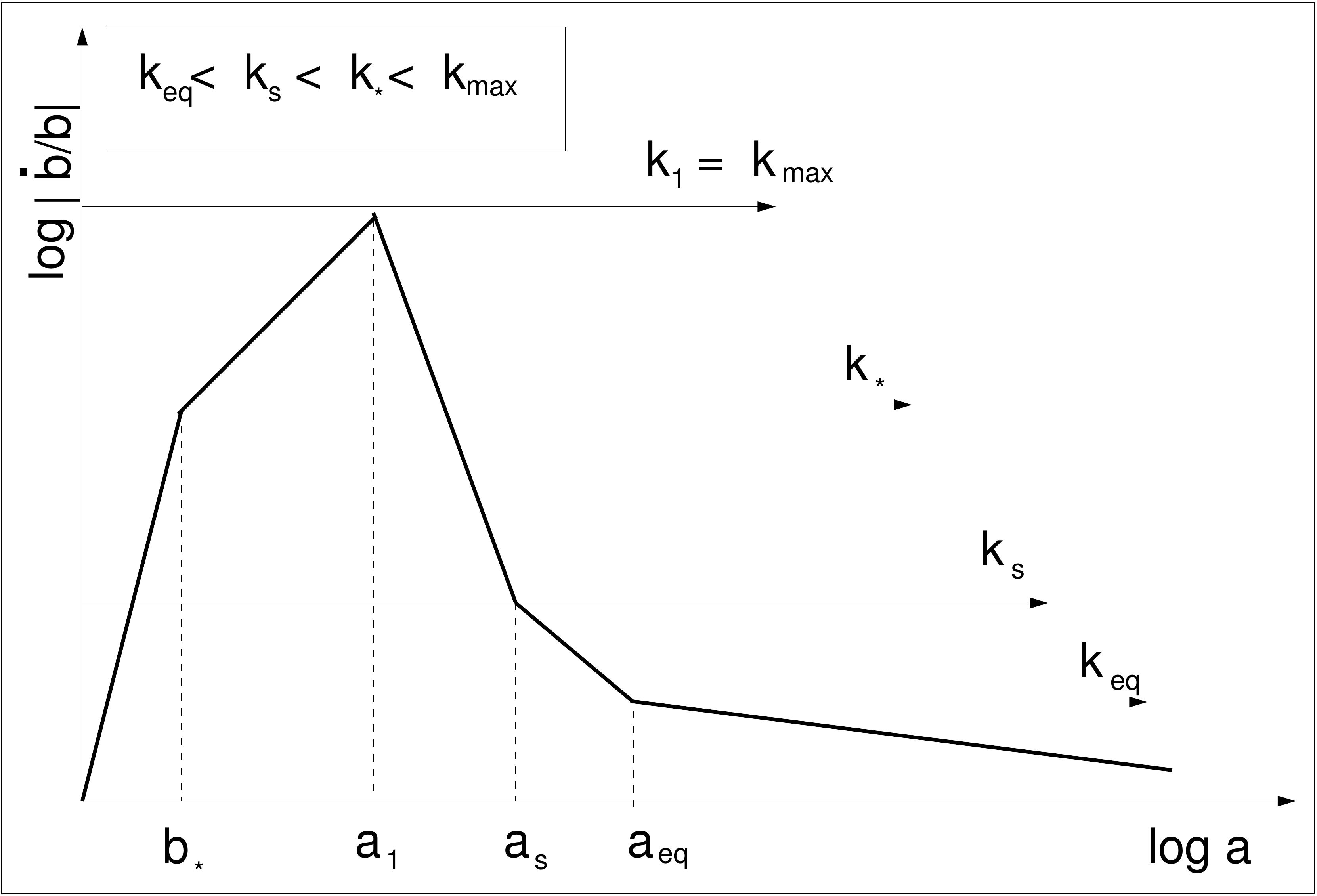}
\caption[a]{The dominance of radiation is delayed but the refractive and the stiff phases are both longer than in  Fig. \ref{FIGU2}.}
\label{FIGU3}      
\end{figure}
In this connection Fig. \ref{FIGU3}  illustrates a possibility complementary to the one 
of Fig. \ref{FIGU2} but leading to a different spectrum. Both in Figs. \ref{FIGU2} and \ref{FIGU3}
a stiff phase precedes the ordinary radiation epoch. However  the refractive and the stiff phases of Fig. \ref{FIGU3} are longer then in Fig. \ref{FIGU2}. 
 This occurrence implies the possibility of modes exiting the effective horizon during the 
 refractive phase and reentering during the stiff phase.  This different dynamical situation implies that the 
intermediate branch of the spectrum (i.e. $a_{*} H_{*} < k < a_{s} H_{s}$) is not quasi-flat anymore. 
Furthermore using Eqs. (\ref{sq1}) and (\ref{sq3}), $\Omega_{gw}$ scales as $|k \tau_{s}|^{s_{T}}$ 
where now $s_{T}$ is given by 
\begin{equation}
s_{T} = 2 - \frac{4}{3 w+1} + n_{T} \simeq 2 -  \frac{4}{3 w+1}  +\frac{\alpha ( 3 - 2 \gamma)}{1 + \alpha} + {\mathcal O}(\epsilon).
\label{sq10}
\end{equation}
For instance, for $w=1$ we will have that $s_{T} = 1 + \alpha$ while $m_{T} = {\mathcal O}(1)$.
The slope of Eq. (\ref{sq10}) is always increasing and since also the other branches 
of the spectral energy density are increasing (thought at a different rate)  all the energy 
of this spectrum will be concentrated in the highest frequency regime and this 
is the reason why the detectability prospects are, in this situation, less promising than in the case of a sufficiently 
long plateau at high frequency. Various other examples can be analyzed by using the 
approximate methods described in this section but they are not central to the present discussion.
\renewcommand{\theequation}{5.\arabic{equation}}
\setcounter{equation}{0}
\section{Detectability prospects}
\label{sec5}
\subsection{Basic considerations}
The phenomenological signatures of the relic gravitons are customarily assessed by using the comoving frequency that is defined as $\nu=k/(2\pi)$ where $k$ denotes the comoving wavenumber. 
Four complementary quantities can be used to describe the cosmic graviton background: 
{\it i)} the tensor power spectrum (denoted by ${\mathcal P}_{T}(\nu, \tau_{0})$),
{\it ii)} the spectral energy distribution (i.e. $\Omega_{gw}(\nu,\tau_{0})$), {\it iii)} the chirp amplitude 
 $h_{c}(\nu,\tau_{0})$ and {\it iv)} the spectral amplitude $S_{h}(\nu,\tau_{0})$ (measured in units of $\mathrm{Hz}^{-1}= \mathrm{sec}$).
Except for $S_{h}(\nu,\tau_{0})$ the three remaining variables are dimensionless. 
The chirp amplitude is, by definition, $h_{c}^2(\nu,\tau_{0}) = {\mathcal P}_{T}(\nu,\tau_{0})/2$. 
The tensor power spectrum at the present time can be related to the spectral energy distribution as 
\begin{equation} 
\Omega_{gw}(\nu,\tau_{0}) = \frac{3 \pi^2 \nu^2}{4 H_{0}^2 a_{0}^2}  {\mathcal P}_{T}(\nu,\tau_{0}).
\label{B1}
\end{equation}
Equation (\ref{B1}) as well as all other equations in this section involve wavelengths 
shorter than the Hubble radius at the present time $\tau_{0}$.
The chirp amplitude and the spectral amplitude are directly related to  $\Omega_{gw}(\nu,\tau_{0})$ in the following manner:
\begin{equation}
h_{c}(\nu,\tau_{0}) = \frac{1}{\pi}\sqrt{\frac{3}{2}} \biggl(\frac{H_{0} a_{0}}{ \nu}\biggr) \sqrt{\Omega_{gw}(\nu,\tau_{0})},\qquad 
{\mathcal S}_{h}(\nu,\tau_{0}) = \frac{3 H_{0}^2 a_{0}^2}{4 \pi^2 \nu^3} \Omega_{gw}(\nu,\tau_{0}).
\label{B3} 
\end{equation}
Equation (\ref{B3}) implies that $h_{c}^2(\nu,\tau_{0})= 2 \nu {\mathcal S}_{h}(\nu,\tau_{0})$ and since 
the detectors of gravitational radiation are operating in the audio band (i.e. between few Hz
and $10$ kHz) it is useful to stress the explicit relations between 
the various quantities mentioned above for typical frequencies $\nu = {\mathcal O}(100)$ Hz 
where the sensitivities of wide-band detectors to cosmic graviton background 
are (approximately) maximal\footnote{Since $\Omega_{gw}$ contains the inverse of $\rho_{crit}$, $h_{0}^2 \Omega_{gw}$ is in fact independent 
on $h_{0}$. } :
\begin{eqnarray} 
h_{c}(\nu,\tau_{0}) &=& 1.263 \times 10^{-20} \biggl(\frac{100 \,\, \mathrm{Hz}}{\nu}\biggr) \, \sqrt{h_{0}^2 \,\Omega_{gw}(\nu,\tau_{0})},
\label{B4}\\ 
{\mathcal P}_{T}(\nu,\tau_{0}) &=& 3.190 \times 10^{-40} \,\,\biggl(\frac{100\,\mathrm{Hz}}{\nu}\biggr)^2 \,\, h_{0}^2 \,\Omega_{gw}(\nu,\tau_{0}),
\label{B5}\\
{\mathcal S}_{h}(\nu,\tau_{0}) &=&  7.981\times 10^{-43} \,\,\biggl(\frac{100\,\mathrm{Hz}}{\nu}\biggr)^3 \,\, h_{0}^2 \,\Omega_{gw}(\nu,\tau_{0})\,\, \mathrm{Hz}^{-1}.
\label{B6}
\end{eqnarray}
From Eqs. (\ref{B4}), (\ref{B5}) and (\ref{B6}) the orders of magnitude of the different  variables 
employed in the description of relic graviton backgrounds 
can be explicitly assessed.
\subsection{Pivotal frequencies}
The spectral energy distribution is characterized by various typical frequencies which are determined from
the wavenumbers appearing in Figs. \ref{FIGU1}, \ref{FIGU2} and \ref{FIGU3}. The smallest frequency range 
of the spectrum follows from the pivot wavenumber $k_{p}$ at which the scalar and tensor power 
spectra are assigned \cite{BICPL,first,wmap9a,wmap9b}:
\begin{equation}
\nu_{p} = \frac{k_{p}}{2\pi} = 3.092\times 10^{-18} \mathrm{Hz} = 3.092 \,\, \mathrm{aHz}.
\label{B7}
\end{equation}
The frequency associated with the dominance of dark energy is of the same order of Eq. (\ref{B7}) 
and it is fixed by $\Omega_{M0}$ and $\Omega_{\Lambda}$; in the case 
of the concordance paradigm we have 
\begin{equation}
 \nu_{\Lambda} = 1.638 \biggl(\frac{h_{0}}{0.719}\biggr) \biggl(\frac{\Omega_{\mathrm{M}0}}{0.258}\biggr)^{1/3} \biggl(\frac{\Omega_{\Lambda}}{0.742}\biggr)^{-1/3} \,\, \mathrm{aHz}.
\label{B7a}
\end{equation}
Since the equality wavenumber is $k_{eq} = 0.0732\,[h_{0}^2 \Omega_{R0}/(4.15\times 10^{-5})]^{-1/2}\,\,  h_{0}^2 \Omega_{M0} \,\, \mathrm{Mpc}^{-1}$  the related frequency $\nu_{eq}$ is:
\begin{equation}
\nu_{eq}  =  1.317 \times 10^{-17} \biggl(\frac{h_{0}^2 \Omega_{\mathrm{M}0}}{0.1364}\biggr) \biggl(\frac{h_{0}^2 \Omega_{\mathrm{R}0}}{4.15 \times 10^{-5}}\biggr)^{-1/2}\,\, \mathrm{Hz}.
\label{B8}
\end{equation}
The frequency $\nu_{bbn} = {\mathcal O}(10^{-2})$ nHz enters directly the 
big-bang nucleosynthesis constraint (see below Eq. (\ref{BBN1})) and sets the scale for the suppression 
of the cosmic graviton background due to neutrino free-streaming \cite{w1,w2}. The explicit 
expression of  the big-bang nucleosynthesis frequency is\footnote{Note that $g_{\rho}$ denotes the effective number 
of relativistic degrees of freedom entering the total energy density of the plasma and $T_{bbn}$ is the putative temperature of big-bang nucleosynthesis. }:
\begin{equation}
\nu_{bbn}= 2.252\times 10^{-11} \biggl(\frac{g_{\rho}}{10.75}\biggr)^{1/4} \biggl(\frac{T_{bbn}}{\,\,\mathrm{MeV}}\biggr) 
\biggl(\frac{h_{0}^2 \Omega_{\mathrm{R}0}}{4.15 \times 10^{-5}}\biggr)^{1/4}\,\,\mathrm{Hz}.
\label{B9}
\end{equation}
The presence of a refractive phase illustrated in Figs. \ref{FIGU1} and \ref{FIGU2} introduces two 
further frequencies:
\begin{eqnarray}
\nu_{*} &=& p(\alpha, \epsilon, N_{*}, N_{t}) \, \nu_{\mathrm{max}}, \qquad  
p(\alpha, \epsilon, N_{*}, N_{t})= \biggl| 1 + \frac{\alpha}{1- \epsilon}\biggr| e^{N_{*}(\alpha+ 1) - N_{t}},
\label{ONEa}\\
\nu_{\mathrm{max}} &=& 1.95\times 10^{8} \biggl(\frac{\epsilon}{0.001}\biggr)^{1/4} 
\biggl(\frac{{\mathcal A}_{{\mathcal R}}}{2.41\times 10^{-9}}\biggr)^{1/4} 
\biggl(\frac{h_{0}^2 \Omega_{R0}}{4.15 \times 10^{-5}}\biggr)^{1/4}  \,\, \mathrm{Hz},
\label{ONEb}
\end{eqnarray}
where ${\mathcal A}_{{\mathcal R}}$ denotes the amplitude 
of the power spectrum of curvature inhomogeneities at the wavenumber $k_{p}$.
Even if Eq. (\ref{ONEb}) suggests that $\nu_{\mathrm{max}} = {\mathcal O}(200)$ MHz,
the value of the end-point frequency of the spectrum may exceed $\nu_{\mathrm{max}}$ since it depends on the post-inflationary 
thermal history \cite{two}. For the thermal histories of Figs. \ref{FIGU2} and \ref{FIGU3}
the spectral energy distribution may extend up to $\nu_{spike} = \nu_{\mathrm{max}}/\sigma > \nu_{\mathrm{max}}$ (with $\sigma<1$).
While a similar spike (with different physical features) may also appear when the refractive index is not dynamical 
\cite{two}, in this particular case two new frequencies appear and they are defined as
\begin{equation}
\nu_{s} =  \sigma^{3(w+1)/(3w -1)} \nu_{\mathrm{max}}, \qquad \nu_{spike} = \nu_{\mathrm{max}}/\sigma, \qquad \sigma =
\biggl(\frac{H_{\mathrm{max}}}{H_{r}} \biggr)^{\frac{1-3w}{6 (w+1)}},
\label{spike}
\end{equation}
where $H_{r}$ denotes the Hubble rate at the onset of the radiation dominance, i.e. right after the stiff phase.
The difference between $\nu_{\mathrm{max}}$ and $\nu_{spike}$ comes essentially from the redshift 
during the stiff stage of expansion. 

\subsection{Phenomenological constraints} 
In the low-frequency range the tensor to scalar ratio of Eq. (\ref{PS7}) 
is bounded from above not to conflict with the observed temperature and polarization 
anisotropies of the CMB; 
in the present analysis we specifically required $r_{T}(\nu_{p}) <0.06$, as it follows 
from a joint analysis of Planck and BICEP2/Keck array data \cite{BICPL}. 
As already mentioned in the introduction slightly less restrictive bounds are 
often used in the current literature and they amount to demanding $r_{T}(\nu_{p}) < {\mathcal O}(0.1)$ 
\cite{wmap9a,wmap9b,planck}. The pulsar timing measurements impose instead the limit $\Omega_{gw}(\nu_{pulsar},\tau_{0}) < 1.9\times10^{-8}$ at the frequency  $\nu_{pulsar} = {\mathcal O}(10) \,\mathrm{nHz}$ corresponding to the inverse of the observation 
time along which the pulsars timing has been monitored \cite{PUL1,PUL2,PUL3,PUL4,PUL5,PUL6}. 
The big-bang nucleosynthesis sets an indirect constraint  
on the extra-relativistic species (and, among others, on the relic gravitons) at the time when light nuclei 
have been formed \cite{bbn1,bbn2,bbn3}. This limit is often expressed in terms of $\Delta N_{\nu}$ 
representing the contribution of supplementary (massless) neutrino 
species (see e.g. \cite{bbn4}) but the extra-relativistic species do not need to be fermionic. 
If, as in our case, the additional species are relic gravitons we will have to demand that:
\begin{equation}
h_{0}^2  \int_{\nu_{bbn}}^{\nu_{\mathrm{max}}}
  \Omega_{gw}(\nu,\tau_{0}) d\ln{\nu} = 5.61 \times 10^{-6} \Delta N_{\nu} 
  \biggl(\frac{h_{0}^2 \Omega_{\gamma0}}{2.47 \times 10^{-5}}\biggr).
\label{BBN1}
\end{equation}
The bounds on $\Delta N_{\nu}$ range from $\Delta N_{\nu} \leq 0.2$ 
to $\Delta N_{\nu} \leq 1$ so that the right hand side of Eq. (\ref{BBN1}) 
turns out to be between $10^{-6}$ and $10^{-5}$.  The basic 
considerations discussed here can be complemented by other bounds which are, however,  less constraining 
than the ones mentioned above. The same logic employed for the derivation of Eq. (\ref{BBN1}) can be applied 
at the decoupling of matter and radiation. While the typical frequency of BBN is ${\mathcal O}(10^{-10})$ Hz the typical frequencies 
of matter radiation equality is ${\mathcal O}(10^{-16})$ Hz (see Eqs. (\ref{B8}) and (\ref{B9})). Since the decoupling between matter and radiation occurs after equality we have that 
\begin{equation}
h_{0}^2  \int_{\nu_{dec}}^{\nu_{max}}
  \Omega_{{\rm GW}}(\nu,\tau_{0}) d\ln{\nu} \leq 8.7 \times 10^{-6}.
 \label{CMBconst} 
  \end{equation}
While the bound itself is numerically similar to the one of Eq. (\ref{BBN1}) the lower extremum of integration is 
smaller since $\nu_{dec} \ll \nu_{bbn}$ (see Eqs. (\ref{B8}) and (\ref{B9})). The bound (\ref{CMBconst}) (discussed in Ref. \cite{cmbcost} with slightly 
different notations) has been also taken into account in the present analysis. However, since we are dealing here with growing spectral energy distributions, Eq. (\ref{CMBconst}) is less constraining: for the same (increasing) slope the lower extremum of integration of Eq. (\ref{CMBconst}) gives a smaller contribution than the one of Eq. (\ref{BBN1}).

\subsection{Spectral energy distribution} 
The analytic estimates of the spectral 
energy density of section \ref{sec4} lead to approximate expressions of the spectral energy distribution; however
 for a more quantitative assessment the cosmic graviton spectrum 
should be expressed in terms of $T_{eq}(\nu, \nu_{eq})$, $T_{*}(\nu, \nu_{*})$ and $T_{s}(\nu, \nu_{s})$ denoting, respectively, 
the transfer functions of the energy density at low, intermediate and high frequencies:
 \begin{eqnarray}
T_{eq}(\nu, \nu_{eq}) &=& \sqrt{1 + c_{eq}\biggl(\frac{\nu_{\mathrm{eq}}}{\nu}\biggr) + b_{eq}\biggl(\frac{\nu_{\mathrm{eq}}}{\nu}\biggr)^2},\qquad c_{eq}= 0.5238,\qquad b_{eq}=0.3537,
\label{Teq}\\
T_{*}(\nu, \nu_{*}) &=& \biggl[1 + c_{*}\biggl(\frac{\nu}{\nu_{*}}\biggr)^{2\epsilon +  n_{T}} + b_{*}\biggl(\frac{\nu}{\nu_{*}}\biggr)^{4 \epsilon + 2 n_{T}}\biggr]^{-1/2},
\qquad c_{*} = b_{*} = {\mathcal O}(1), 
\label{Tstar}\\
T_{s}(\nu, \nu_{s}) &=& \sqrt{ 1 + c_{s}  \biggl(\frac{\nu}{\nu_{s}}\biggr)^{p(w)/2} + b_{s}  \biggl(\frac{\nu}{\nu_{s}}\biggr)^{p(w)}}, \qquad 
p(w) = 2 - \frac{4}{3w +1},
\label{Ts}
\end{eqnarray}
where the subscripts refer to the typical frequencies involved in each transition, 
i.e. $\nu_{eq}$, $\nu_{*}$ and $\nu_{s}$. To transfer the spectral energy density inside the Hubble 
radius the procedure is to integrate numerically the equations of the tensor modes; the derivation of $T_{eq}(\nu,\nu_{eq})$ and $T_{s}(\nu,\nu_{s})$, in a different physical situation, has been discussed in detail in \cite{absolute,N1,N2}.  In the literature it is also customary to introduce 
the transfer function of the power spectrum \cite{sptr1,sptr2} and the two transfer functions have slightly different numerical features 
that have been discussed in the past (see e.g. Ref. \cite{absolute} for a comparison). With these specifications, we have:
 \begin{eqnarray}
h_{0}^2 \,\Omega_{gw}(\nu,\tau_{0}) &=& {\mathcal N}_{\rho} \,\, r_{T}( \nu_{p})\,\, {\mathcal T}^2(\nu, \nu_{eq}, \nu_{*}, \nu_{s}) \, \,\biggl(\frac{\nu}{\nu_{\mathrm{p}}} \biggr)^{n_{\mathrm{T}}} \, e^{- 2 \,\beta\,\nu/\nu_{\mathrm{max}}}, 
\label{OMA}\\
{\mathcal T}(\nu, \nu_{eq}, \nu_{*}, \nu_{s}) &=& T_{eq}(\nu, \nu_{eq})\, T_{*}(\nu, \nu_{*}) \, T_{s}(\nu, \nu_{s}),
\label{OMB}\\
{\mathcal N}_{\rho} &=& 4.165 \times 10^{-15}\, \biggl(\frac{h_{0}^2 \Omega_{\mathrm{R}0}}{4.15\times 10^{-5}}\biggr) \, \biggl(\frac{{\mathcal A}_{{\mathcal R}}}{2.41\times 10^{-9}}\biggr), 
\label{OMC}
\end{eqnarray}
where $n_{T} = [\alpha ( 3 - 2 \gamma) - 2 \epsilon]/(1 + \alpha - \epsilon)$ (see also Eq. (\ref{sq8a})) and $r_{T}(\nu_{p})$ is the tensor to scalar ratio of Eq. (\ref{PS7}) evaluated at the pivot frequency $\nu_{p}$. In the conventional case $r_{T}(\nu_{p})$ is related to the slow-roll parameter $\epsilon$  and to the tensor spectral index $n_{T}$ via the so-called consistency relations (see also, in this respect, Ref. \cite{violation} where a similar model 
for the violation of the consistency relations has been discussed). In the present situation $r_{T}(\nu)$ and $n_{T}$ {\em do not} obey the 
consistency relations and depend on the rate of variation of the refractive index $\alpha$ and on the critical number of efolds $N_{*}$. If the refractive index is not dynamical (i.e. $\alpha \to 0$ and $\gamma \to 0$) we have, as expected, that $n_{T} \to - 2 \epsilon$. In Eq. (\ref{OMA}) $\beta$ is a parameters  ${\mathcal O}(1)$ which depends upon the width of the transition between the inflationary phase and the subsequent radiation dominated phase;  for different widths of the post-inflationary transition we can estimate $0.5 \leq \beta \leq 6.3$ \cite{N1,N2}. 
The numerical coefficients appearing in Eqs. (\ref{Teq}), (\ref{Tstar}) and (\ref{Ts}) are determined from each 
specific transition: while $c_{eq}$ and $b_{eq}$ can be accurately assessed, $c_{*}$ and $b_{*}$ depend 
on the parametrization of the refractive index and, similarly $c_{s}$ and $b_{s}$ change depending on the values of $w$.
In the case $w\to 1$ there are even logarithmic corrections which have been specifically scrutinized in the past\footnote{According to Eq. (\ref{Teq}), $T_{eq}(\nu) \to 1$ for $\nu \gg \nu_{eq}$ but the realistic situations further suppressions are expected. 
The neutrino free-streaming produces an effective anisotropic stress leading ultimately to an integro-differential 
equation (see, for instance, \cite{w1,w2}). This aspect will be discussed later when assessing the other minor sources of damping.} (see e.g. \cite{two}).

\subsection{The constrained parameter space}
We shall be predominantly
interested in the possibility of a relatively strong signal in the audio and 
in the mHz bands. We remind that the audio band ranges between few Hz and $10$ kHz where the terrestrial wide-band 
interferometers operate. The mHz band ranges instead between a fraction of the mHz and the Hz; in this 
range space-borne detectors might one day hopefully within the following score year. The MHz band, extending between
 $100$ kHz and few GHz; this band is immaterial for potential signals coming from conventional inflationary models but could play 
a relevant role in the present context since, in this case, most of the signal is concentrated exactly 
in this region.
\begin{figure}[!ht]
\centering
\includegraphics[height=7.2cm]{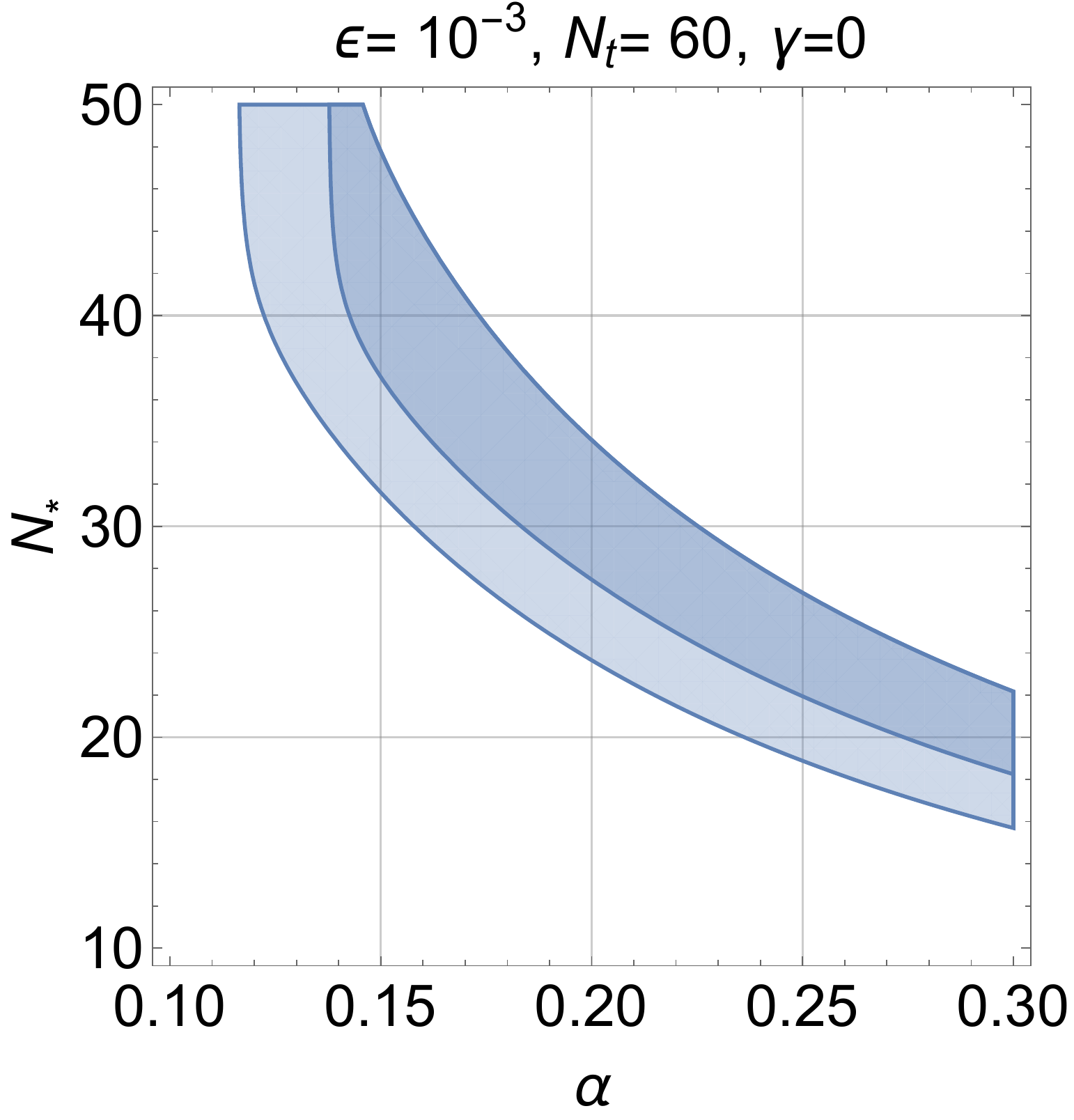}
\includegraphics[height=7.2cm]{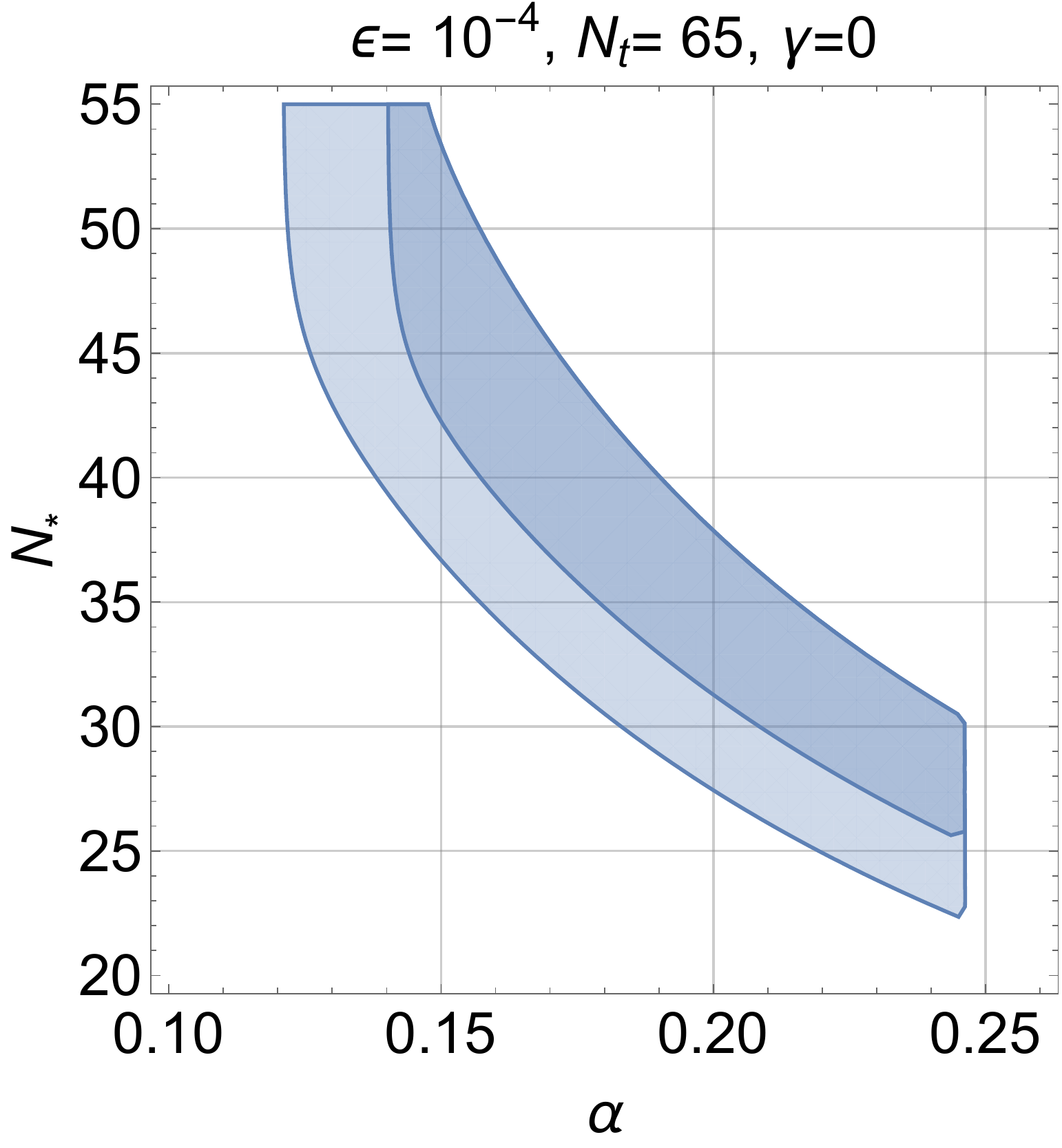}
\caption[a]{The outer an the inner shaded areas illustrate the regions of the parameter space where 
all the constraints are satisfied in conjunction either with the requirements of Eqs. (\ref{CON3})--(\ref{CON4}) 
(outer regions) or with the stronger demands of Eqs. (\ref{CON5})--(\ref{CON6}) (inner regions). Both plots refer to the case $\gamma=0$. }
\label{FIGU4}      
\end{figure}
To account for the possibility of a detection in the audio band we shall then impose 
on the parameter space a further constraint on the chirp amplitude:
\begin{equation} 
h_{c}(\nu_{audio}, \tau_{0}) > 10^{-25}, \qquad \nu_{audio} = 0.1 \,\, \mathrm{kHz},
\label{CON1}
\end{equation}
where $\nu_{audio}$ roughly corresponds, in practice, to the expected maximum of the sensitivity 
for the (advanced) Ligo/Virgo intereferometers. In the mHz band we shall instead 
require: 
\begin{equation} 
h_{c}(\nu_{mHz}, \tau_{0}) > 2\times 10^{-23}, \qquad \nu_{mHz} = \mathrm{mHz}.
\label{CON2}
\end{equation}
Equations (\ref{CON1}) and (\ref{CON2}) imply that 
we should select regions of the parameter space where the spectral 
energy distribution exceeds, respectively, $10^{-11}$ and $10^{-16}$;
more specifically we are led to demand:
\begin{eqnarray}
&& h_{0}^{2} \Omega_{gw}(\nu_{audio}, \tau_{0}) >  6.2 \times 10^{-11}, \qquad \nu_{audio}= 0.1\,\,\mathrm{kHz},
\label{CON3}\\
&&  h_{0}^{2} \Omega_{gw}(\nu_{mHz}, \tau_{0}) >  2.5 \times 10^{-16},\qquad \nu_{mHz} = \mathrm{mHz}.
\label{CON4}
\end{eqnarray}
Since these requirements might not be achieved with rushing speed, we shall also 
consider a couple of less pretentious conditions, namely 
\begin{eqnarray}
&& h_{0}^{2} \Omega_{gw}(\nu_{audio}, \tau_{0}) >  10^{-9}, \qquad \nu_{audio}= 0.1\,\,\mathrm{kHz},
\label{CON5}\\
&&  h_{0}^{2} \Omega_{gw}(\nu_{mHz}, \tau_{0}) >   10^{-12},\qquad \nu_{mHz} = \mathrm{mHz}.
\label{CON6}
\end{eqnarray}
While from the viewpoint of the experiments Eqs. (\ref{CON5})--(\ref{CON6}) are weaker than Eqs. (\ref{CON3})--(\ref{CON4}), 
from the viewpoint of the signal itself the opposite is true: if we enforce Eqs. (\ref{CON3})--(\ref{CON4}) the allowed 
region of the parameter space will be larger than in the case of Eqs. (\ref{CON5})--(\ref{CON6}). See, in this 
respect, Figs. \ref{FIGU4} and \ref{FIGU5} where we illustrate the constrained parameter 
space for different choices of the parameters and in the case 
of a conventional thermal history. 
\begin{figure}[!ht]
\centering
\includegraphics[height=7.2cm]{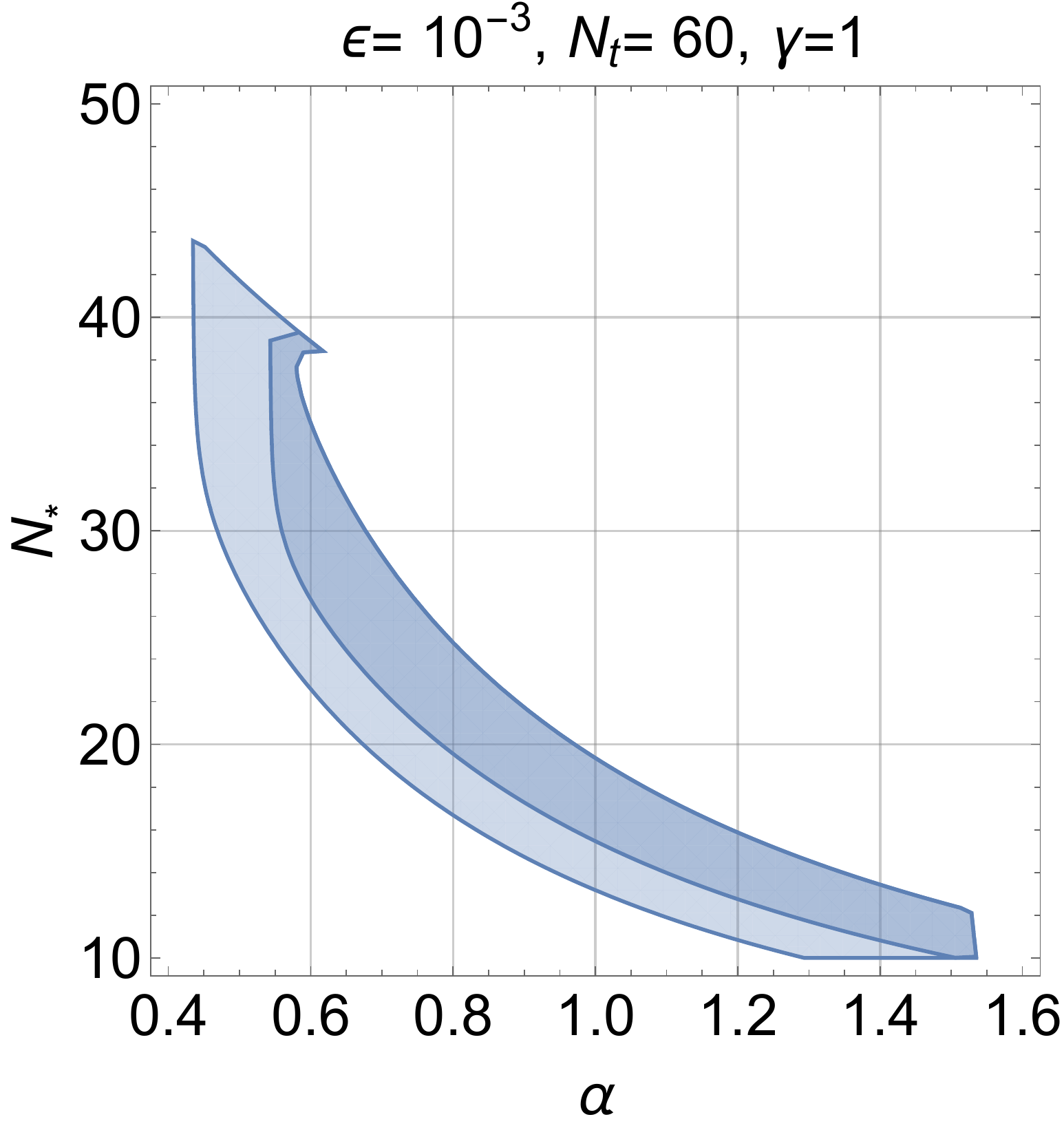}
\includegraphics[height=7.2cm]{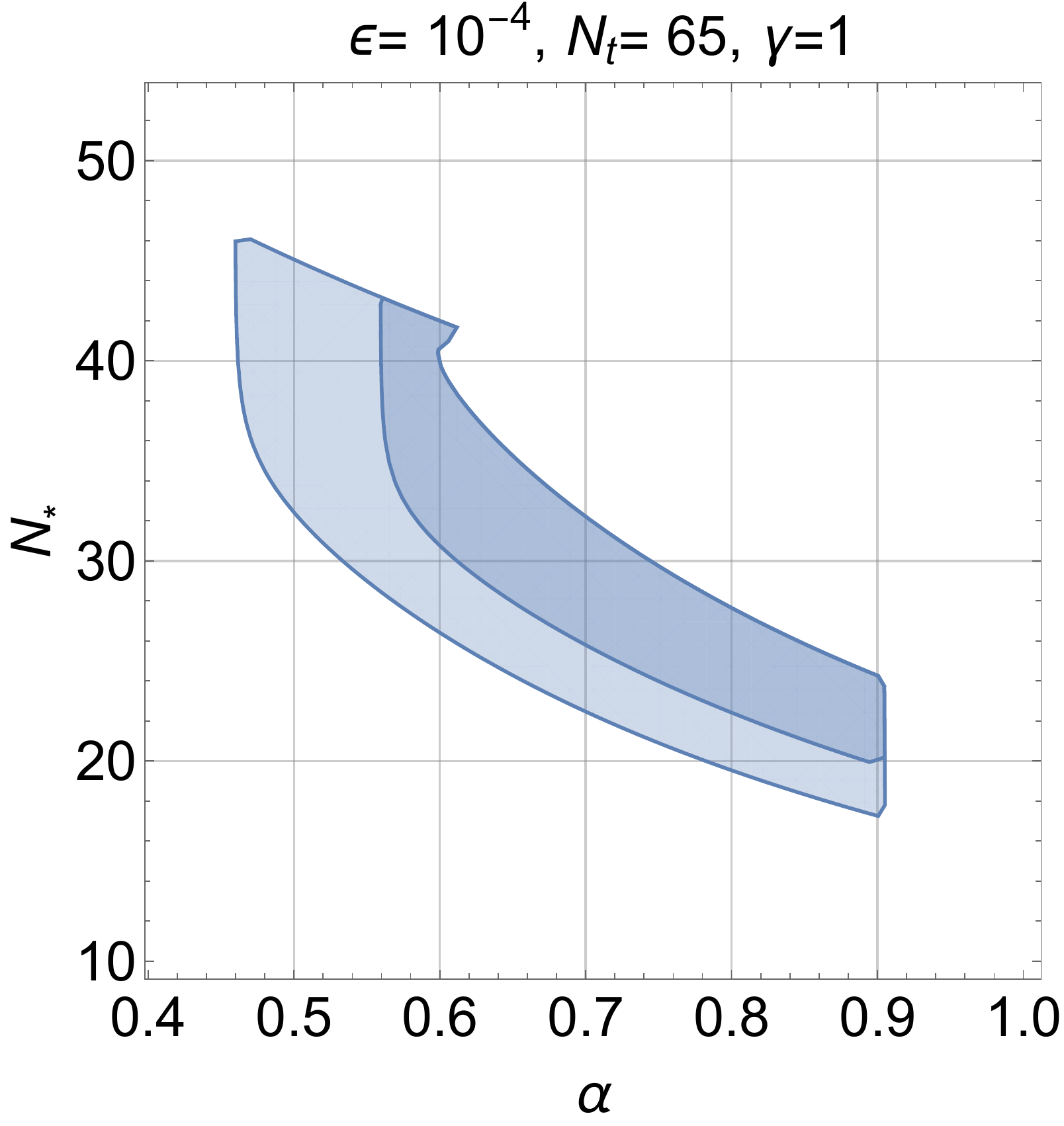}
\caption[a]{The same analysis of Fig. \ref{FIGU4} is illustrated in the case $\gamma = 1$. }
\label{FIGU5}      
\end{figure}
The shaded areas in both plots  describe the regions where 
all the phenomenological constraints are concurrently 
satisfied while the chirp amplitudes are sufficiently large to 
be detected.
More specifically in Figs. \ref{FIGU4} and \ref{FIGU5} the outer regions are obtained by 
enforcing the requirements of Eqs. (\ref{CON5}) and (\ref{CON6}). Conversely the inner 
regions come from the more demanding conditions spelled out in Eqs. (\ref{CON3}) 
and (\ref{CON4}). The reduction of the areas between the outer and the inner 
regions illustrate the reduction of the parameter space induced by the difference 
between the requirements of Eqs. (\ref{CON3})--(\ref{CON4}) and (\ref{CON5})--(\ref{CON6}).
Note that the case $\gamma =0$ is illustrated in Fig. \ref{FIGU4} while Fig. \ref{FIGU5} 
concerns the case $\gamma=1$. Different values of $\gamma$ rescale, in practice, the 
values of $\alpha$, as expected from the general relation connecting $n_{T}$ to $\alpha$ and 
$\gamma$. Indeed, to leading order in $\epsilon$, the value of $n_{T}$ is the case $\gamma \to 0$ 
is roughly thrice its value in the $\gamma \to 1$ case. We therefore have that the range of $\alpha$ 
in the two situations must be rescaled by a factor of $3$ and this is what we clearly 
see by comparing the horizontal axes in Figs. \ref{FIGU4} and \ref{FIGU5}.

\begin{figure}[!ht]
\centering
\includegraphics[height=7.2cm]{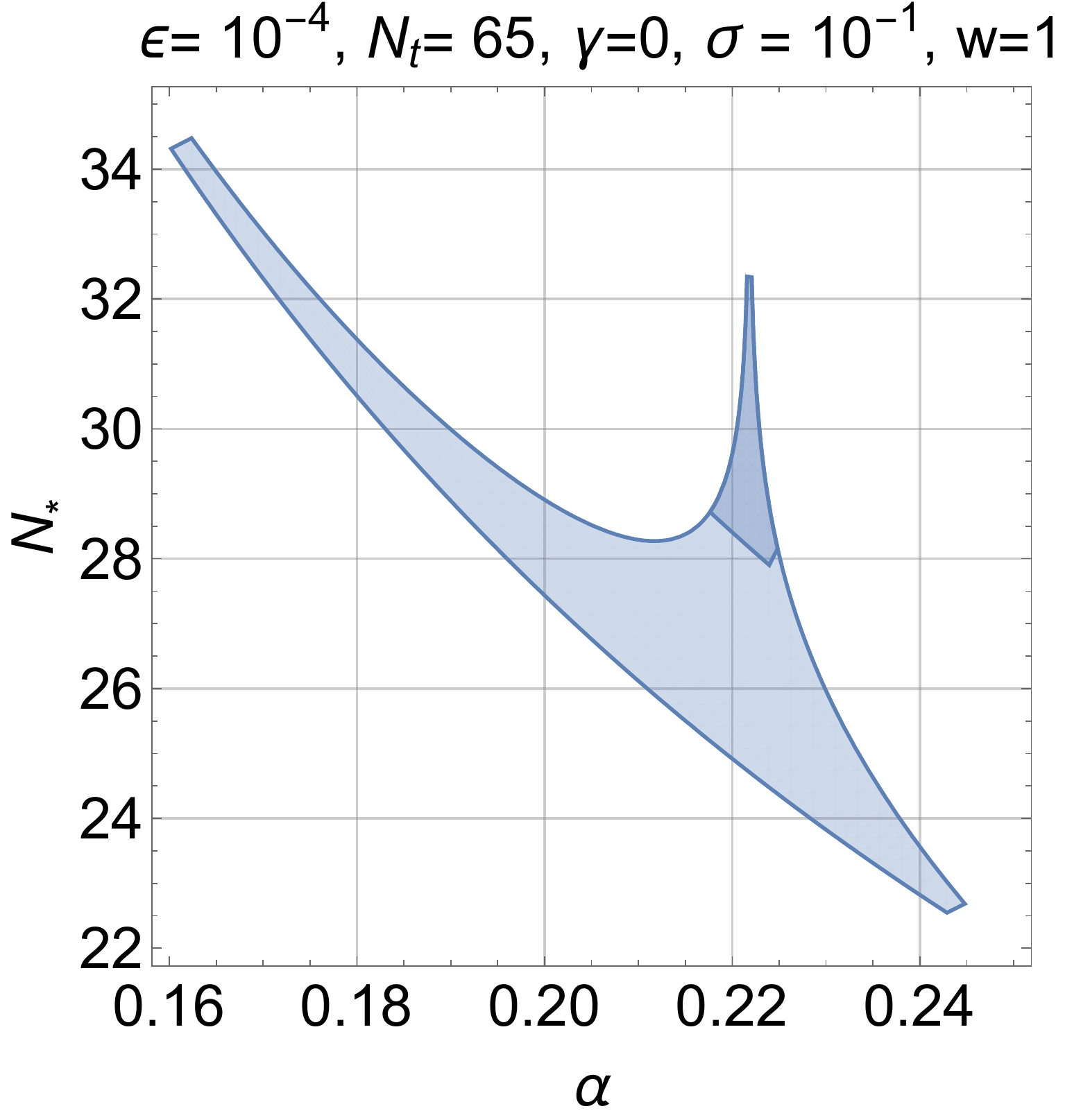}
\includegraphics[height=7.2cm]{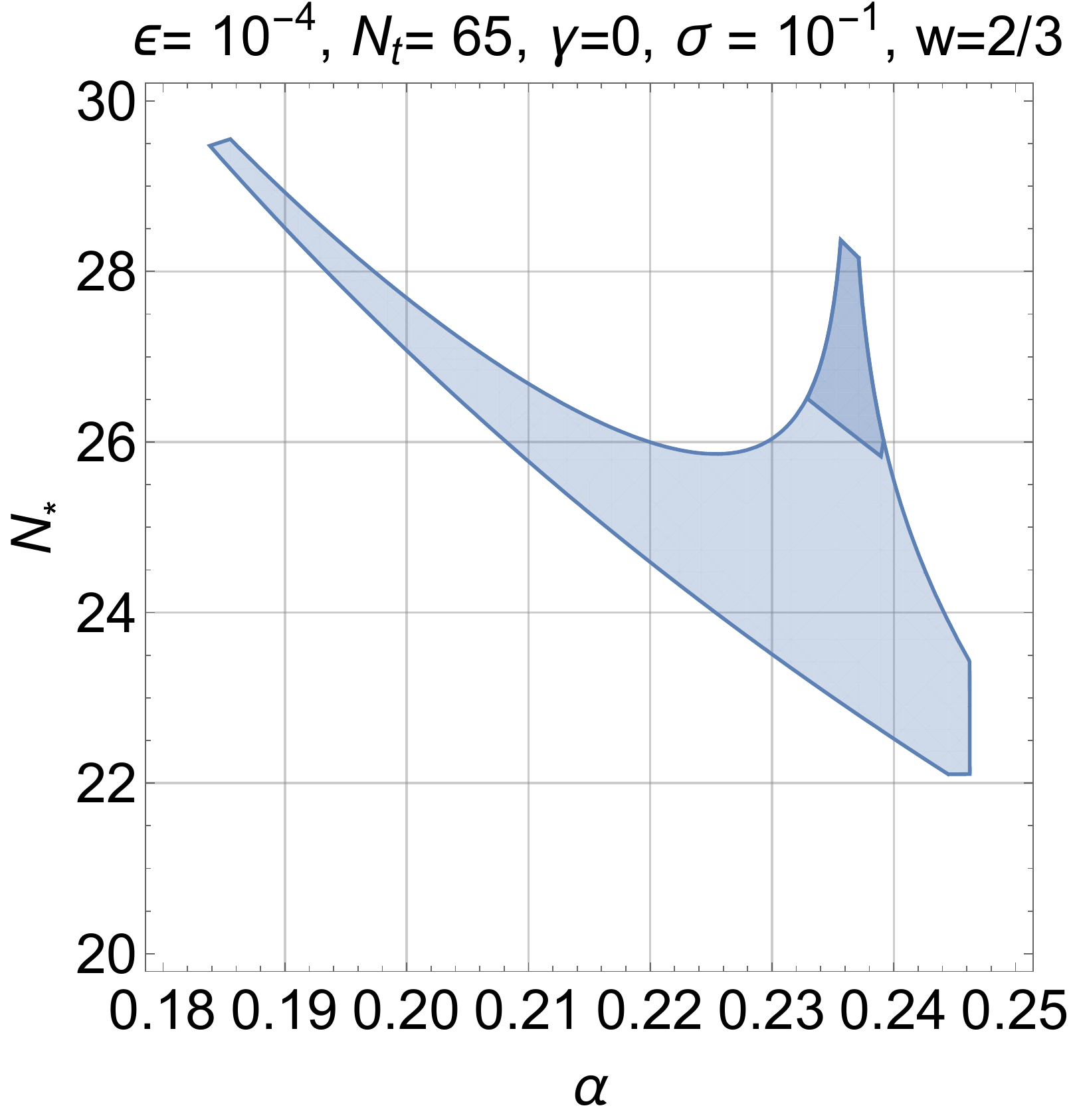}
\includegraphics[height=7.2cm]{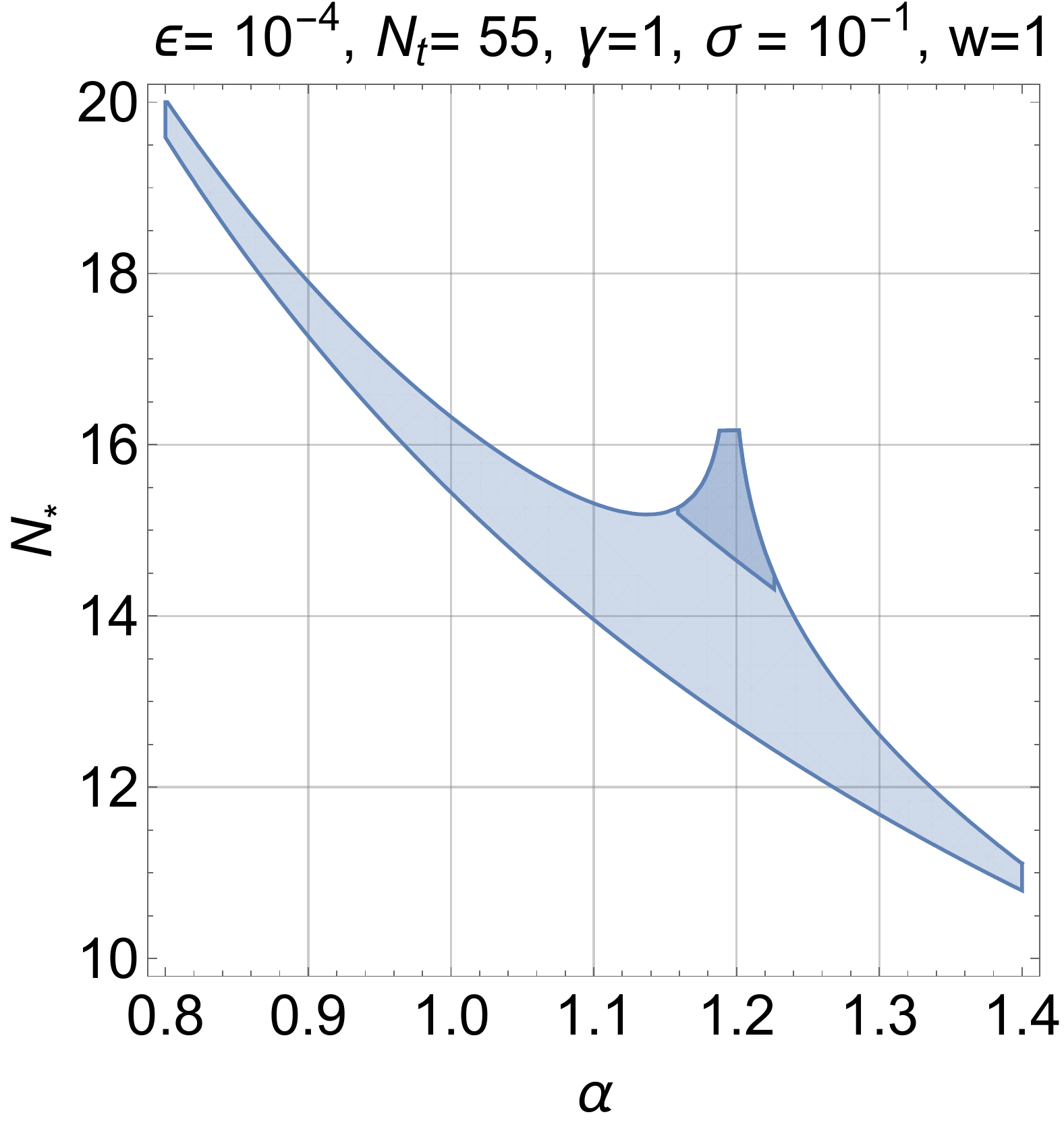}
\includegraphics[height=7.2cm]{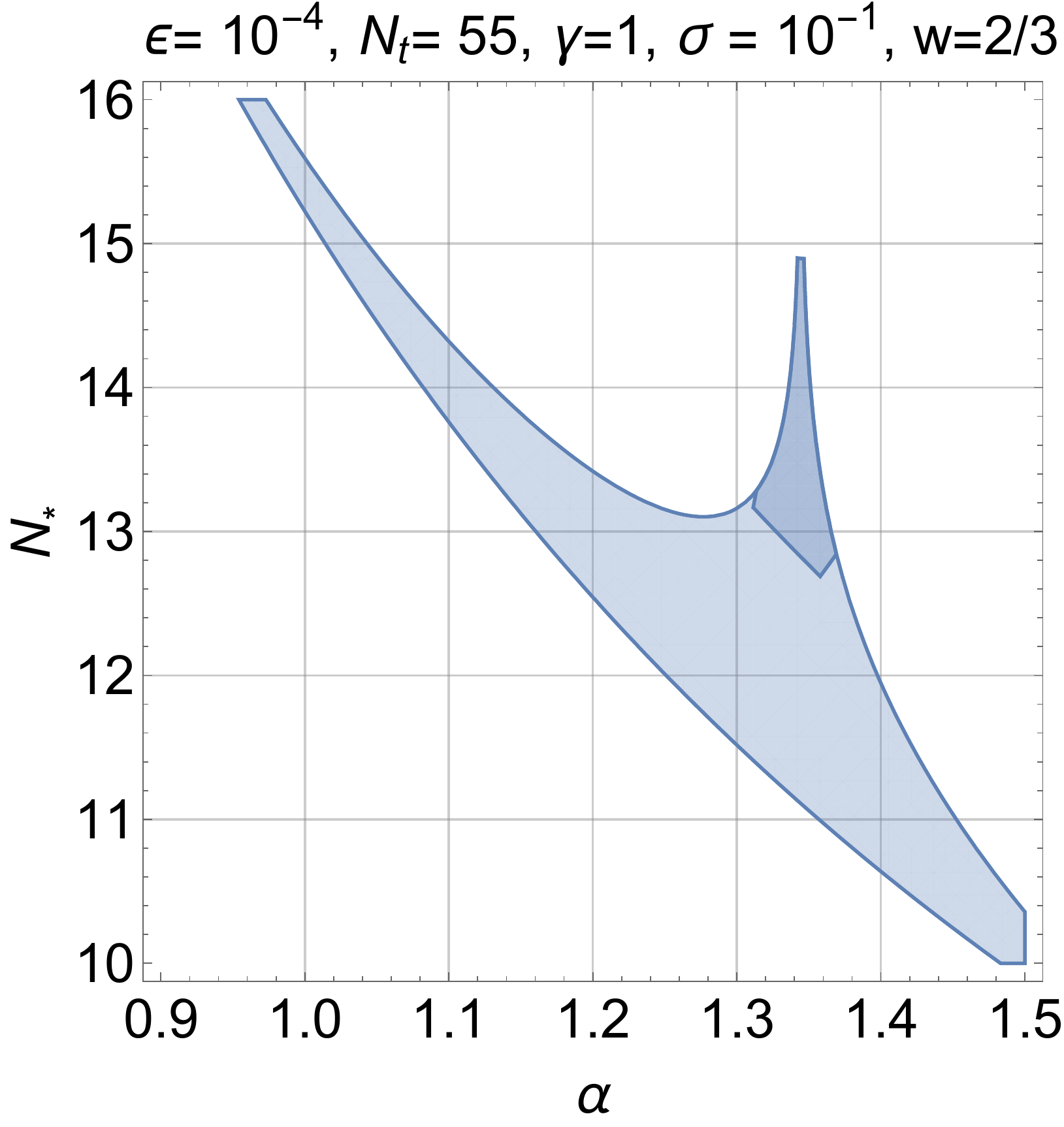}
\caption[a]{We illustrate the constrained parameter space in the 
case of a thermal history characterized by a stiff post-inflationary 
phase. As in the case of Figs. \ref{FIGU4} and \ref{FIGU5} the outer and the inner
regions refer, respectively,  to the requirements of Eqs. (\ref{CON3})--(\ref{CON4}) and (\ref{CON5})--(\ref{CON6}).}
\label{FIGU6}      
\end{figure}

In the case of a different post-inflationary history the constrained parameter 
space gets modified and the relevant exclusion plots are illustrated in Fig. \ref{FIGU6} 
for a fiducial choice of the parameters. In the two plots at the right 
the barotropic index corresponds to $2/3$ while in the two plots 
at the left the barotropic index is maximal (i.e. $w=1$). 
By looking at the inner and at the outer exclusion regions we conclude 
that a reduction in the sensitivities of the hypothetical detectors 
drastically reduces the areas of the parameter space. Indeed, as in Figs. \ref{FIGU4} and 
\ref{FIGU5} the inner and the outer plots correspond, respectively, to the requirements of 
Eqs. (\ref{CON5})--(\ref{CON6}) and to the requirements of Eqs. (\ref{CON3})--(\ref{CON4}).
The reason for this reduction is a direct consequence of the violet spectral slope 
in the highest frequency domain. Still, for a given value of $\sigma$, the 
constrained parameter space suggests a potentially interesting signal.

The explicit profiles of different models will now be illustrated. While the same 
analysis can be easily rephrased either in term of the power spectrum ${\mathcal P}_{T}(\nu,\tau_{0})$ 
or in terms of the spectral amplitude $S_{h}(\nu,\tau_{0})$ we shall be mainly interested in the chirp amplitude and in the spectral energy distribution. This kind of approach is also useful for the explicit derivation of a template family.
\begin{figure}[!ht]
\centering
\includegraphics[height=10.4cm]{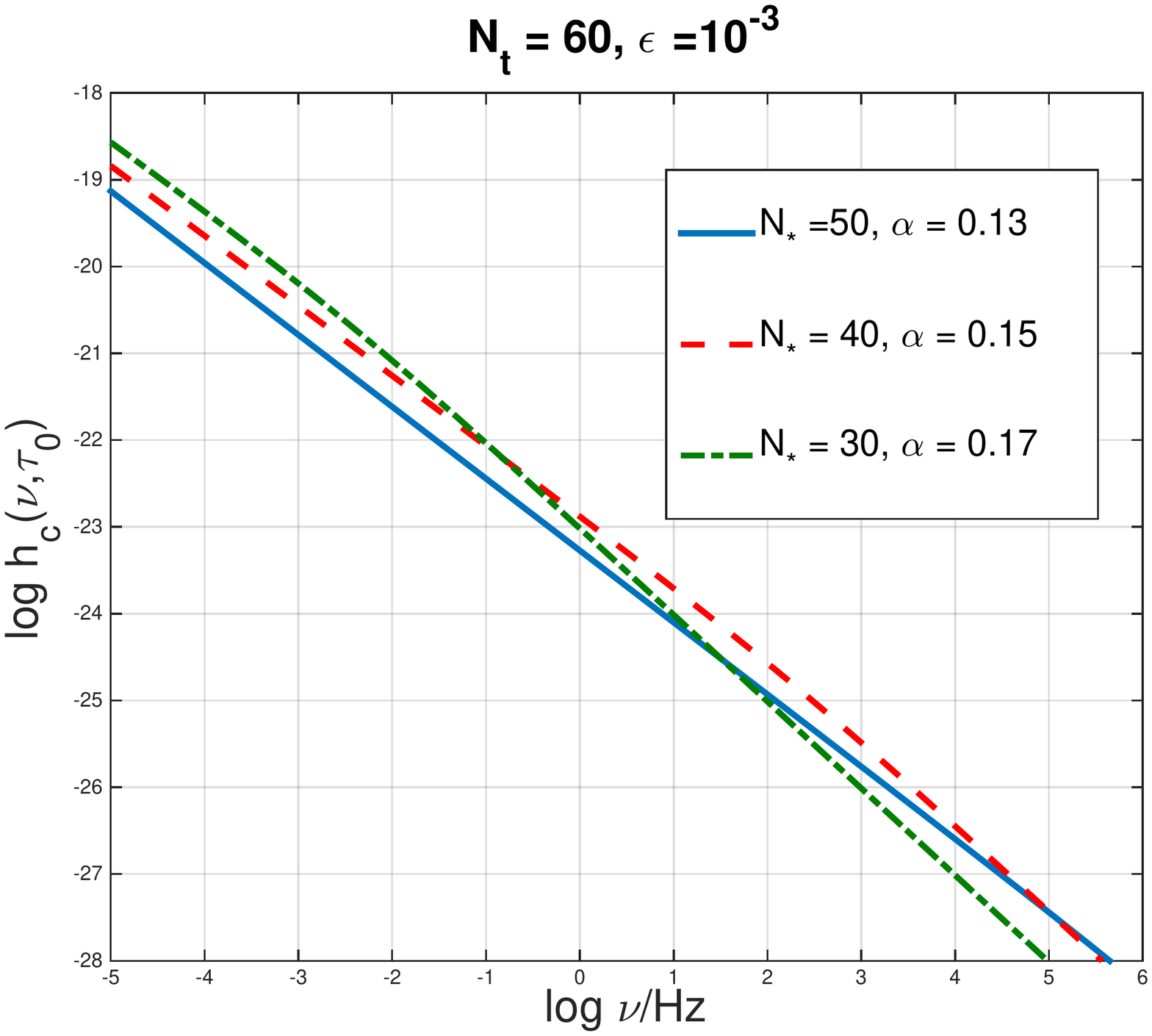}
\includegraphics[height=10.4cm]{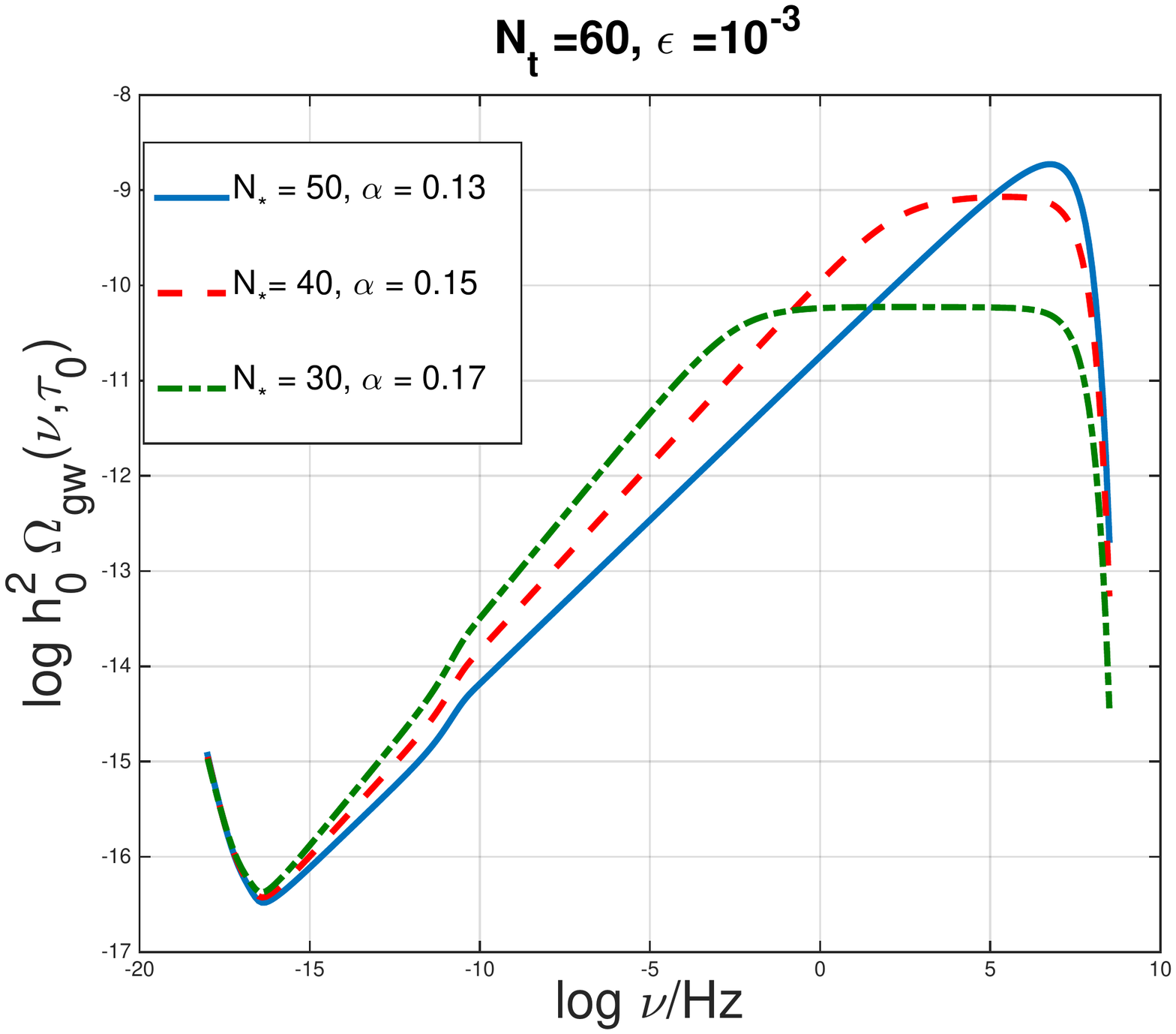}
\caption[a]{The chirp amplitude and the spectral energy distribution produced 
by a dynamical refractive index
are illustrated in the case of a standard post-inflationary thermal history.  Note that in this 
and in the following two figures we took $\gamma = 0$. }
\label{FIGU7}      
\end{figure}

\begin{figure}[!ht]
\centering
\includegraphics[height=10.6cm]{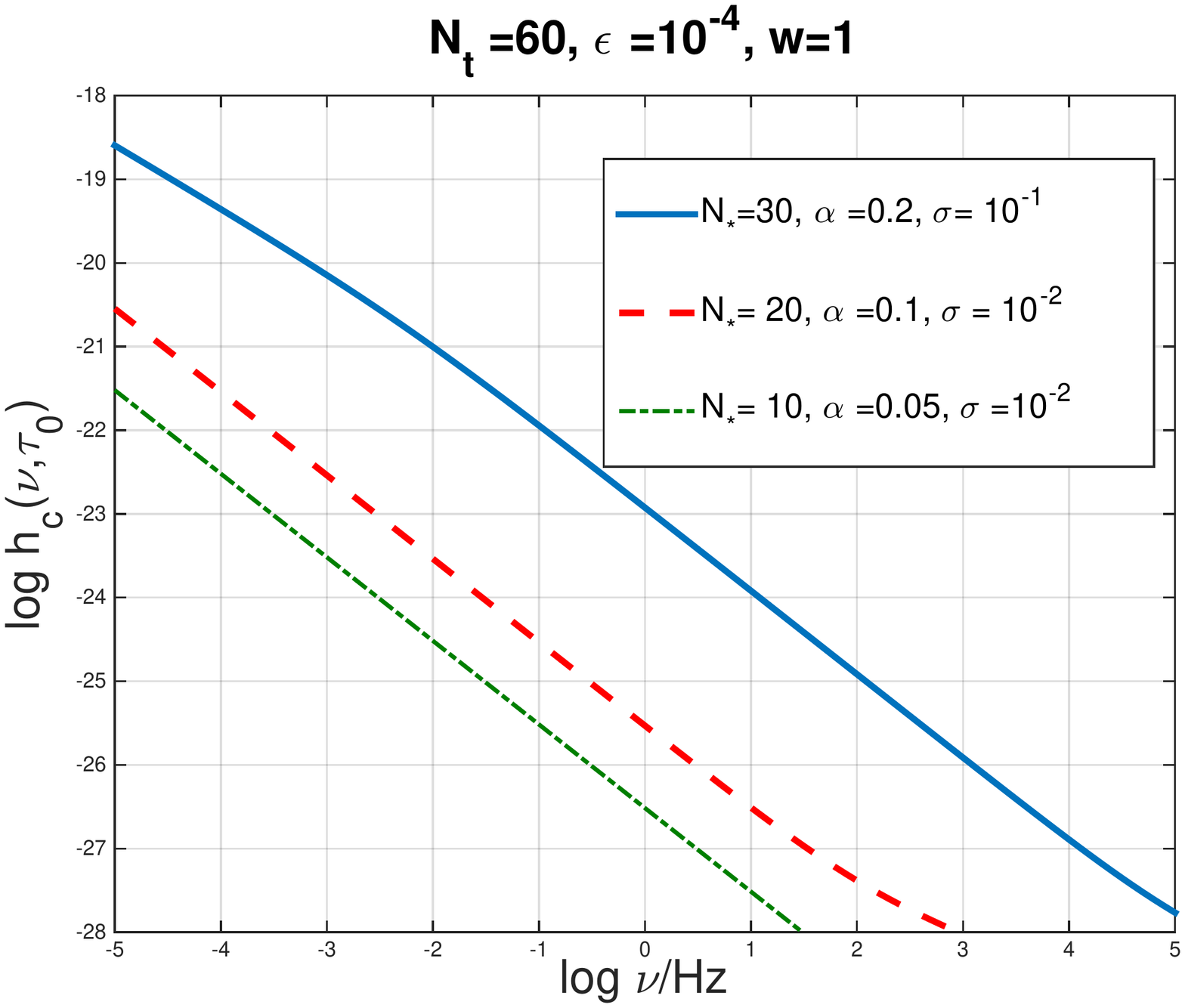}
\includegraphics[height=10.6cm]{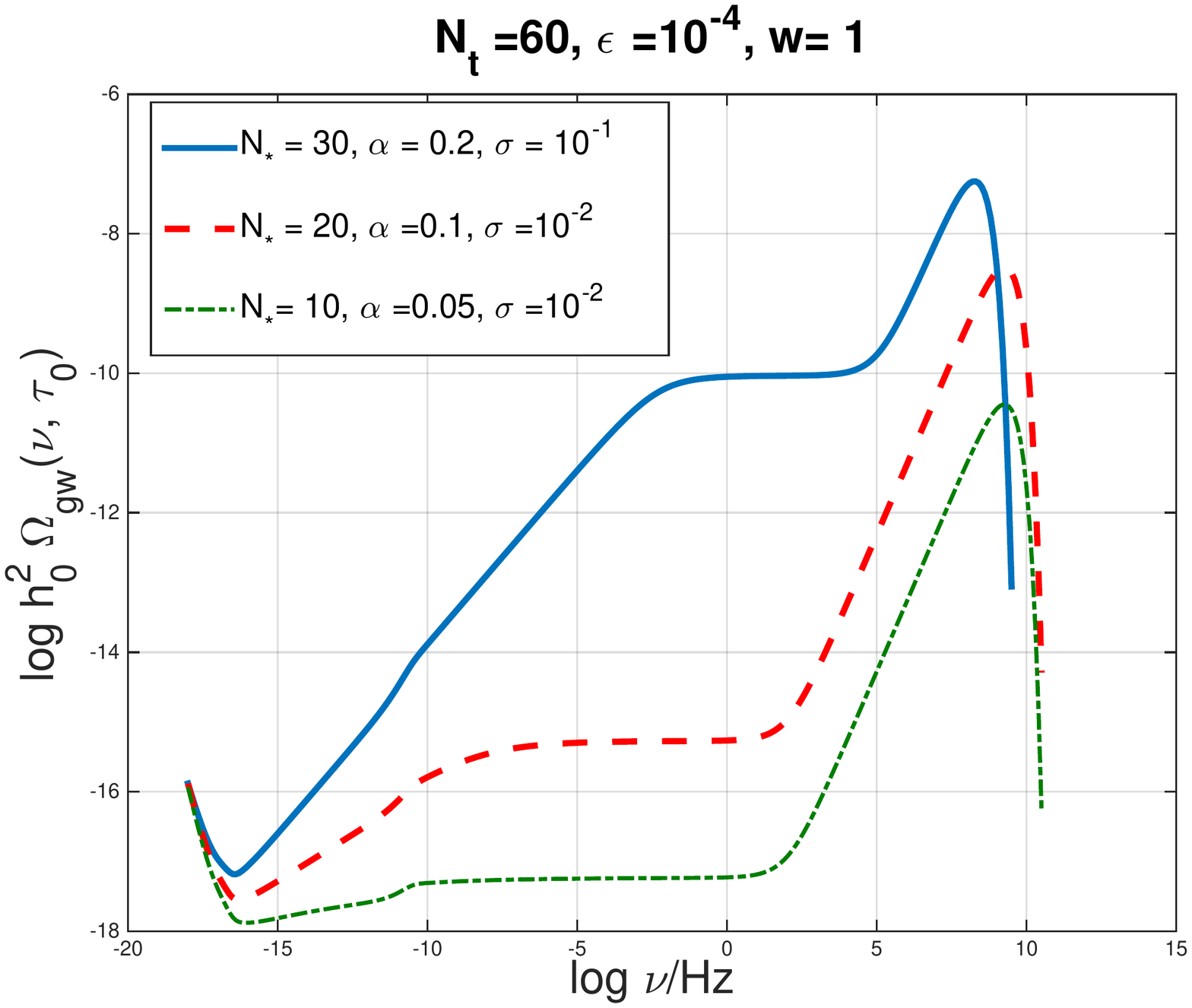}
\caption[a]{The relic graviton background is illustrated in the case of a dynamical refractive index 
and assuming a post-inflationary thermal history that includes a stiff phase with $w=1$. }
\label{FIGU8}      
\end{figure}

\begin{figure}[!ht]
\centering
\includegraphics[height=10.6cm]{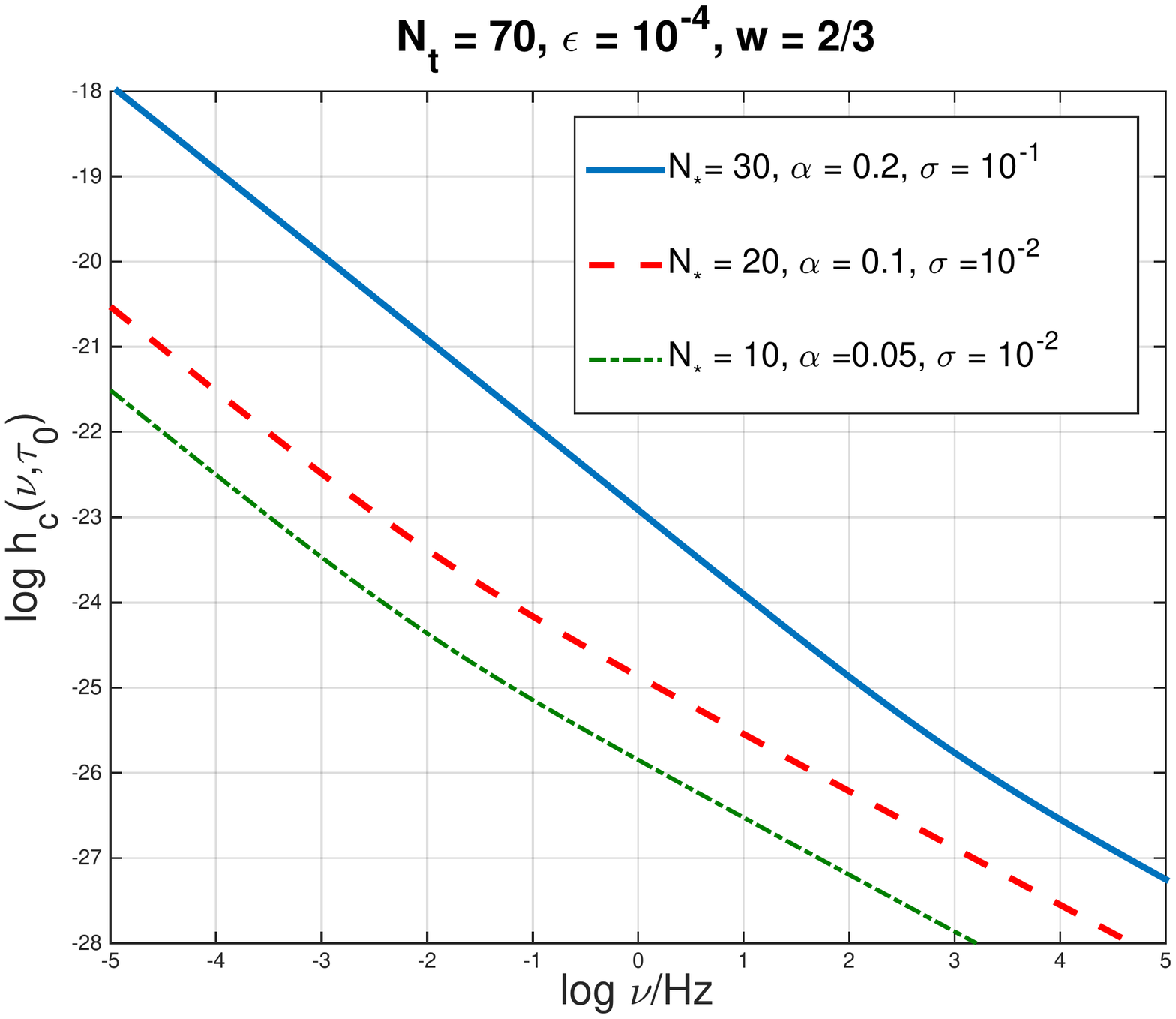}
\includegraphics[height=10.6cm]{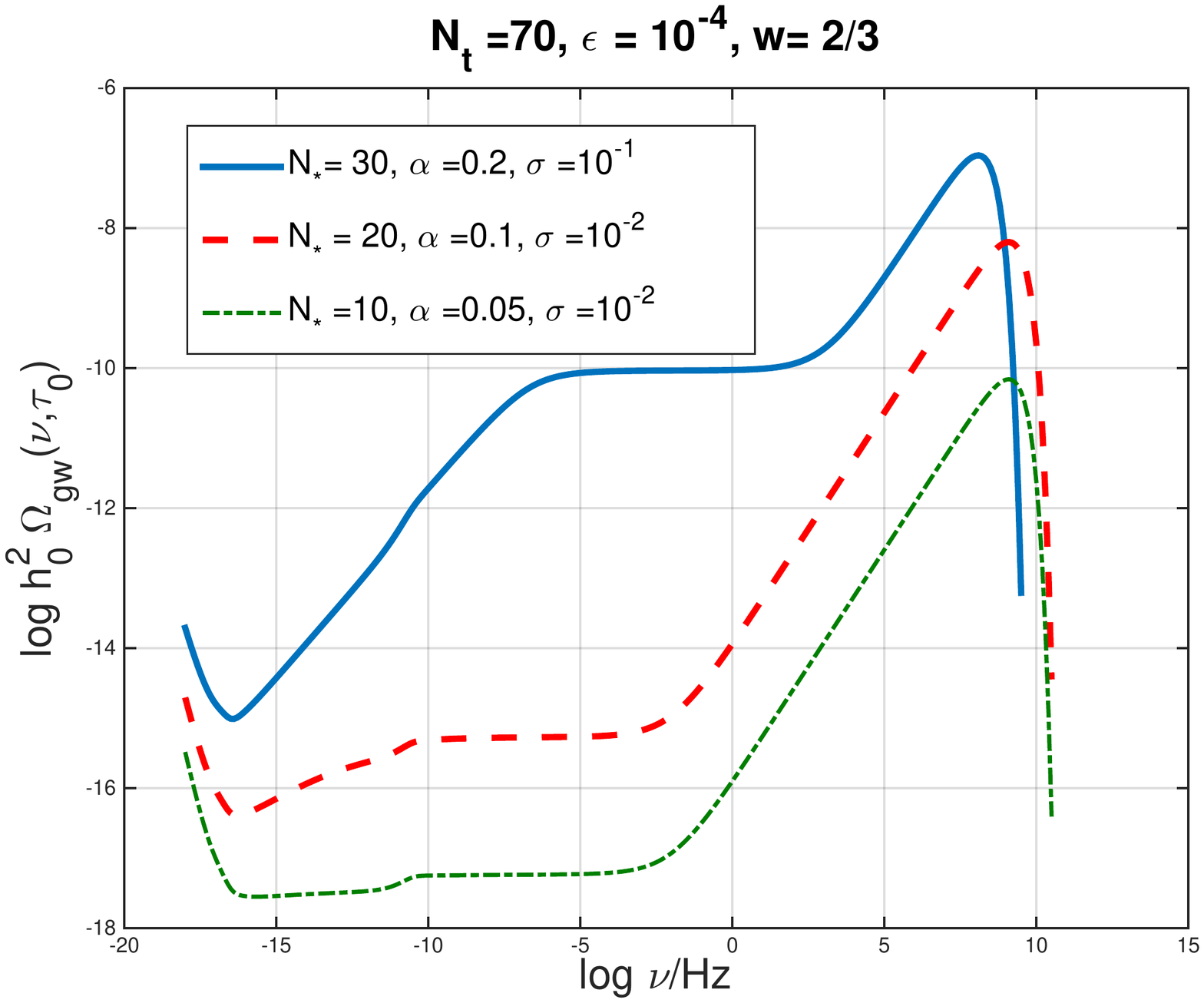}
\caption[a]{The same analysis of Fig. \ref{FIGU8} is  repeated in the case $w = 2/3$ for slightly 
different values of the parameters to illustrate potential degeneracies in the parameter 
space. }
\label{FIGU9}      
\end{figure}

In Fig. \ref{FIGU7} we illustrate the chirp amplitude and the spectral energy 
distribution in the case of a standard post-inflationary history. 
The scales of the two plots on the horizontal axis are 
different\footnote{ The chirp amplitude is illustrated in a frequency range 
encompassing the mHz and the audio bands. Conversely 
the spectral energy distribution covers all the frequencies 
from the aHz up to the MHz band.}. The explicit 
differences among the various models are less 
pronounced if we look at the chirp amplitude which is 
is proportional to the square root of the spectral 
energy distribution and it is further suppressed 
by one power of the frequency. 
It is finally relevant to stress that Figs. \ref{FIGU7}, \ref{FIGU8} and \ref{FIGU9} 
have been all derived in the case $\gamma =0$. This is not a limitation, as 
we saw from the discussion of the constrained parameter space: different values 
of $\gamma$ simply shift the allowed range of $\alpha$. As $N_{*}$ increases 
the plateau of Fig. \ref{FIGU7} becomes less evident.
The results of Fig. \ref{FIGU7} can be usefully compared with the ones illustrated in Fig. 
\ref{FIGU8} where the post-inflationary evolution is characterized by a stiff epoch.
While in Fig. \ref{FIGU8} we considered the case $w=1$, in Fig. \ref{FIGU9} we took instead 
$w = 2/3$.
Recalling Eqs. (\ref{sq9})--(\ref{sq10})  and (\ref{spike}) the value of the barotropic index 
controls not only the slope of the cosmic graviton spectrum in the vicinity high-frequency spike
but also the frequency range. Larger values of $w$ correspond to more violet slopes in the MHz
band while the values of $\sigma$ are inversely proportional to the frequency of the spike so that 
as $\sigma$ gets smaller than $1$ the position of the spike exceeds the GHz. Both $\sigma$ and $w$ 
determine the length of the stiff phase. 

In the right plots of Figs. \ref{FIGU7}, \ref{FIGU8} and \ref{FIGU9} we can appreciate a minor suppression for frequencies of the order of $0.01$ nHz. This suppression is less visible in the chirp amplitude but it is evident from the spectral energy distribution. For $\nu< \nu_{bbn}$ the slight break in the spectrum is due to the neutrino free 
streaming. The neutrinos free stream, after their decoupling, and the effective energy-momentum tensor acquires, to first-order in the amplitude 
of the plasma fluctuations, an anisotropic stress.  The overall effect of collisionless particles is a reduction 
of the spectral energy density of the relic gravitons. Assuming that the only collisionless 
species in the thermal history of the Universe are the neutrinos, the amount 
of suppression can be parametrized by the function
\begin{equation}
{\mathcal F}(R_{\nu}) = 1 -0.539 R_{\nu} + 0.134 R_{\nu}^2, \qquad R_{\nu} = \frac{r}{r + 1}, \qquad r = 0.681 \biggl(\frac{N_{\nu}}{3}\biggr),
\label{ANIS3}
\end{equation}
where, as usual,  $R_{\nu}$ is the fraction of neutrinos in the radiation plasma; clearly in the concordance 
model $R_{\gamma} + R_{\nu} =1 $. In the case $R_{\nu}=0$ (i.e. in the absence of collisionless particles) there is no suppression. If, on the contrary, 
$R_{\nu} \neq 0$ the suppression can even reach one order of magnitude. In the case $N_{\nu} = 3$, 
$R_{\nu} = 0.405$ and the suppression of the spectral energy density is proportional 
to ${\mathcal F}^2(0.405)= 0. 645$.  This suppression due to neutrino free streaming is thus effective for frequencies larger than $\nu_{\mathrm{eq}}$ and smaller than $\nu_{bbn}$.

Besides neutrino free streaming there are other two minor effects taken into account in Figs. \ref{FIGU7}, \ref{FIGU8} and \ref{FIGU9}: the damping effect associated with the (present) dominance of the dark energy and the suppression due to the variation of the effective number of relativistic species.  In the concordance scenario the redshift of $\Lambda$-dominance  (i.e. $(\Omega_{\Lambda}/\Omega_{M0})^{1/3}$) determines the numerical 
value of $\nu_{\Lambda}$ defined in Eq. (\ref{B7a}). The adiabatic damping of the mode function due to the dominance of the dark energy implies a damping of the order of $(\Omega_{M0}/\Omega_{\Lambda})^2$ in the spectral energy distribution. This suppression competes with a potential increase of the spectral energy distribution for $\nu < \nu_{\lambda}$ and going as $(\nu/\nu_{\Lambda})^{-2}$ \cite{l1,zh}. 
Finally, for temperatures much larger than the top quark mass, all the known species of the minimal standard model of particle interactions are in local thermal equilibrium, then $g_{\rho} = g_{\mathrm{s}} = 106.75$. Below $T \simeq 175$ GeV the various species start decoupling, the notion of thermal equilibrium is replaced by the notion of kinetic equilibrium and the time evolution of the number of relativistic degrees of freedom effectively changes the evolution of the Hubble rate. 
In principle if a given mode $k$ reenters the Hubble radius at a temperature $T_{k}$ the spectral energy density 
of the relic gravitons is (kinematically) suppressed by a factor which can be written as \cite{N1,N2}
$(g_{\rho}(T_{k})/g_{\rho0}) (g_{\mathrm{s}}(T_{k})/g_{\mathrm{s}0})^{-4/3}$ where, at the present time,  $g_{\rho0}= 3.36$ and $g_{\mathrm{s}0}= 3.90$. 
So, in the case of the minimal standard model the suppression on $\Omega_{gw}(\nu,\tau_{0})$
 will be of the order of $0.38$. In popular supersymmetric extensions of the minimal standard models $g_{\rho}$ and $g_{s}$  can be as high as, approximately, $230$. This will bring down the figure given above to $0.29$.
\newpage
\renewcommand{\theequation}{6.\arabic{equation}}
\setcounter{equation}{0}
\section{Concluding remarks}
\label{sec6} 
The evolution of the refractive index during the early stages of a conventional inflationary phase leads 
to a spectral energy distribution naturally tilted towards frequencies and different post-inflationary 
thermal histories typically add a further branch to the cosmic graviton spectrum. 
Overall the spectrum may then exhibit up to four different branches extending between the 
aHz region and the GHz band. Assuming, in a minimalistic perspective, that  the evolution of the 
refractive index terminates before the end of inflation, the spectral energy distribution involves a quasi-flat plateau at high frequencies that is supplemented by a spike between the MHz and the GHz. 
Depending on the thermal history the slopes of the spectral energy distribution are red, 
blue and even violet. After imposing the usual phenomenological constraints, there are 
still wide portions of the parameter space where the resulting signal could be detectable, 
at least in principle, either by terrestrial interferometers (in their advanced and enhanced 
configuration) or by space-borne detectors. In spite of less mundane 
possibilities leading to growing spectral energy distributions of relic gravitons, 
the present findings demonstrate that blue and violet spectra 
are compatible with conventional inflationary scenarios
 in the presence of a dynamical refractive index. 

\section*{Acknowledgements}

I wish to thank M. Doser, F. Fidecaro, M. Pepe-Altarelli, G. Unal and  for interesting exchanges on the topics discussed in this paper. I also thank T. Basaglia, A. Gentil-Beccot and S. Rohr of the CERN Scientific Information Service for their kind assistance.

\end{document}